\documentclass[12pt, draftclsnofoot, onecolumn]{IEEEtran}
\usepackage{amsmath,amsfonts}
\usepackage{algorithmic}
\usepackage{algorithm, color}
\usepackage{array}
\usepackage[caption=false,font=normalsize,labelfont=sf,textfont=sf]{subfig}
\usepackage[font=small]{caption}
\usepackage{textcomp}
\usepackage{stfloats}
\usepackage{url}
\usepackage{verbatim}
\usepackage{graphicx}
\usepackage{cite}
\usepackage[T1]{fontenc}
\usepackage{theorem}
\usepackage{amssymb}

\newtheorem{theorem}{Theorem}\newtheorem{lemma}{Lemma}\theoremheaderfont{\normalfont\bfseries}

\begin{document}

\title{A Framework for Mutual Information-based MIMO Integrated Sensing and Communication Beamforming Design}
\author{Jin Li, Gui Zhou, Tantao Gong, Nan Liu
\thanks{(Corresponding author: Nan Liu) 
	This article was presented in part at the ICCC conference \cite{conference}.
	
	J. Li, T. Gong and N. Liu are with the National Mobile Communications Research Laboratory, Southeast University, Nanjing 211189, China. (e-mail: lijin, gongtantao, nanliu@seu.edu.cn). G. Zhou is with the Institute for Digital Communications, Friedrich-Alexander-University Erlangen-N\"{u}rnberg (FAU), 91054 Erlangen, Germany (email: gui.zhou@fau.de).}}
%
%

\maketitle

\begin{abstract}
\begin{abstract}
	Integrated sensing and communication (ISAC) unifies sensing and communication, and improves the efficiency of the spectrum, energy, and hardware.
	In this work, we investigate the ISAC beamforming design 
	to maximize the mutual information between the target response matrix of a  point radar target and the
	echo signals, while ensuring the data rate requirements of the communication users.  
	In order to study the impact of the echo interference caused by communication users on  sensing performance, we study two scenarios: a single communication user and multiple communication users. For the case of a single communication user, we consider three types of echo interference, no interference, a point interference, and an extended interference. For the case of multiple communication users, the interference is also an extended one, and furthermore, each user's communication rate requirement needs to be satisfied. To find the optimal beamforming design in these problems, we provide a closed-form solution with low complexiy, a semidefinite relaxation (SDR) method, a low-complexity algorithm based on the Majorization-Minimization (MM) method and the successive convex approximation (SCA) method,  and an algorithm based on MM method and SCA method, respectively.
	Numerical results demonstrate that, compared to the ISAC beamforming schemes based on the Cramér-Rao bound (CRB) metric and the beampattern metric, 
	the proposed mutual information metric can bring better beampattern and root mean square error (RMSE) of angle estimation.
	Furthermore, our proposed schemes designed based on the mutual information metric can suppress the echo interference from the communication users effectively.
\end{abstract}
\end{abstract}

\begin{IEEEkeywords}
Integrated sensing and communication (ISAC), beamforming, mutual information, semidefinite relaxation (SDR), Majorization-Minimization (MM).  
\end{IEEEkeywords}
\vspace{-5mm}

\section{Introduction}
With the rapid increase of the number of wireless equipments, the existing spectrum is getting more and more crowded.  Thus, it is important to seek ways to make use of the spectrum band of higher frequency or reuse the existing spectral resources. Currently, radar occupies plentiful spectral band, and if wireless communications can make good use of the radar spectrum, it will allieviate the pressure of limited spectrum resources for wireless communications. Hence, integrated sensing and communication (ISAC) is a promising technology to unify radar systems and wireless communication systems. 
As a paradigm of ISAC, dual-functional radar-communication (DFRC) systems use the same signal for both radar and
communication, which greatly improves the efficiency
of the spectrum, hardware and energy compared with the radar and communication co-existence (RCC) systems \cite{overview}, \cite{fantwc}.

With the deployment of multiple-input multiple-output (MIMO), the DFRC 
base station (BS) can fully exploit the spatial degree of freedoms (DoFs) to enhance the performance of communication and sensing \cite{fantcom}. 
In communication systems, we have many performance metrics such as transmission rate, bit error rate, and symbol error rate. 
Meanwhile, in sensing systems, we study the problem of parameter estimation and target detection. Hence, Cramér-Rao bound (CRB), target detection probability, squared error of beampattern, mean squared error (MSE), and  output signal-to-interference-noise ratio (SINR) are  popular metrics to measure the performance of  parameter estimation and target detection.
Recently, minimizing the squared error between the ideal and practical beampatterns under communication SINR constraint and BS transmit power constraint was proposed in \cite{fantwc}. It shows that with beamforming design, the DFRC BS with shared transmit and receive antennas has better beampattern than DFRC BS with separated transmit and receive antennas. The authors of \cite{CRB} optimized the CRB  for the MIMO DFRC BS, while at the same time achieved 
better root mean square error (RMSE) and beampattern  
than the beampattern matching design in \cite{fantwc}. In addition, the CRB was proved to be equal to the MSE in the case of sensing extended targets \cite{CRB}, \cite{MSE}. However, the metrics based on estimation and detection often have very complicated mathematical expressions which are hard to optimize such as target detection probability in \cite{tdp}, CRB in \cite{MLE}, and MSE in \cite{MSE}. Furthermore, CRB is the theoretical lower bound of unbiased estimator which is difficult to achieve, and target detection probability depends on the specific detector such as Neyman-Pearson detector \cite{kay}. 

Information theoretic measures in radar systems are also popular metrics. They often have concise mathematical expressions and are applicable to all radar systems \cite{im}, \cite{mi2022}. 
Among radar performance metrics based on information theoretic measures, mutual information (MI) is  a very useful design criterion which has many benefits. It has been proved in \cite{mi} that a high mutual information between the target response matrix and the receiving echo signals implies that radar can achieve accurate classification and estimation performance. The authors of \cite{miandmse} have shown that under the same transmit power constraint,  the performance of MIMO radar waveforms obtained by optimizing the mutual information or the minimum mean squared error (MMSE) are the same when the target impulse vector follows the Gaussian distribution. 
In the space-time coding design of MIMO radar, code construction based on the MI coincides with code construction based on the Chernoff bound under the assumption of uncorrelated sensing targets \cite{chernoff}. 
In the RCC system, designing waveforms by maximizing the MI is beneficial to the co-existence of the MIMO radar and the communication in spectrally crowded environments \cite{tangboradar}. For waveform design of adaptive distributed MIMO radar, maximizing the MI could bring better target detection probability and improve delay-Doppler resolution \cite{adaptiveradar}. 

Due to these benefits, in this paper, we adopt the MI between the target response and the receiving echo signals as the radar design criterion  in ISAC systems while 
ensuring the data rate requirements of communication.
More specifically, in this work, we consider a MIMO DFRC BS that works as a colocated MIMO radar \cite{lijian} with compact antenna spacing to sense targets and communicates with users, simultaneously.
We assume a point radar target which is far away from the BS. 
The MI-based ISAC beamforming design is investigated for two scenarios: a single communication user and multiple communication users. 
%
%
%
In the case of a single communication user, we study three problems: 1) no interference. This happens when the communication user is very far way from the BS, and the average strength of the interfering echo signal from the communication user is very weak and thus, the interference can be ignored; 2) a point interference. This is the case where the communication user is relatively far away from the BS or has a small area, so that the interfering echo signal can be regarded as a point scatterer; 3) an extended interference. This happens when the communication user is near to the BS and/or has a large area, such as a vehicle. 
For the first problem of no interference, we provide a closed-form solution for the ISAC beamforming design with low complexity.
The second problem, which corresponds to a point interference, has an untractable objective function and is converted into a rank-1 relaxed semidefinite programming (SDP) problem through a series of mathematical transformations and the application of Schur complement. Then, rank one solution can be reconstructed with the use of the Gaussian randomization method. 
As for the third problem, which is the case where there is an extended interference, we transform this problem into  a second-order cone programming (SOCP) problem with the aid of
the Majorization-Minimization (MM) method.
Then, a suboptimal closed-form solution can be obtained by means of the Lagrangian dual decomposition method. The computational complexity of the proposed algorithm is much less than that of solving an SOCP problem by CVX.
For the scenario of multiple communication users, which as a whole forms an extended interference, an iterative algorithm based on the MM and the successive convex approximation (SCA) method is proposed. 

The contributions and novelties of our paper are summarized as follows:


$\bullet$ 
We adpot the MI as the radar performance metric to design the ISAC beamforming subject to the constraints of the communication rate and BS  transmit power.
While a useful design criterion in radar systems, MI has not been studied well in ISAC systems. 
Previous literature has studied other radar performance metrics such as the CRB \cite{CRB}, beampattern \cite{fantwc, beampattern}, and MSE \cite{MSE}. We find that the MI metric has a concise mathematical expression and is easy to solve compared to some other estimation and detection metrics such as target detection probability, CRB and MSE. Additionally, we show that the ISAC beamforming scheme based on the MI can bring better beampattern and angle estimation performance, and suppress the interfering echo signals effectively. 

$\bullet$
For the case of a single communication user where the interfering echo signal is very weak, though the problem can be solved by the SCA or SDR method, we provide a closed-form solution which enables fast computation up to 0.009s.

$\bullet$
In the scenario of a single communication user who is a point scatterer, to deal with the nonconvexity of the objective function, we apply a series of mathematical transformations, Schur complement and rank one relaxation, and transform the optimization problem into an SDP problem where the rank one solution is obtained by the Gaussian randomization method.  

$\bullet$
As for the case of a single communication user who is an extended echo intereference, with the application of the MM method, the problem can be transformed into an SOCP problem solved using CVX. However, we propose a low-complexity algorithm based on the Lagrangian dual decomposition method. Compared to the results of CVX, our proposed algorithm can achieve the same performance with much less computational complexity.  



The remainder of this paper is organized as follows. In Section \uppercase\expandafter{\romannumeral2}, we provide the system model of the DFRC BS and the design criteria of radar performance and communication performance. In Sections \uppercase\expandafter{\romannumeral3}, \uppercase\expandafter{\romannumeral4}, and \uppercase\expandafter{\romannumeral5}, we investigate three types of radar interference from a single communication user. In Section \uppercase\expandafter{\romannumeral6}, the scenario of multiple communication users is studied. Sections \uppercase\expandafter{\romannumeral7} and \uppercase\expandafter{\romannumeral6}
provide  numerical results and  conclusions, respectively.

\textbf{Notations:} Throughout this paper, we use the following mathematical notations.
Boldface uppercase letters denote matrices, boldface lowercase letters denote vectors, and normal font denotes scalars. Transpose, conjugate, and hermitian are represented by superscript $(\cdot)^T$, $(\cdot)^*$, and $(\cdot)^H$, respectively. 
The pseudo-inverse of a matrix $\mathbf{A}$ is denoted by $\mathbf{A}^{\dagger}$. Complex field is denoted by $\mathbb{C}$, and  real field is denoted by $\mathbb{R}$. 
$\mathbf{A} \otimes \mathbf{B} $ denotes the Kronecker product between matrices $\mathbf{A}$ and $\mathbf{B}$. $\text{vec} (\mathbf{A})$ denotes the vectorization of matrix $\mathbf{A}$ and
$\text{Blkdiag} (\left[ \mathbf{A}, \mathbf{B}\right])$ is the block diagonalization operation.
Complex Gaussian distribution is represented by $\mathcal{CN}$.
The determinant, trace, and real part of a matrix are represented by $\det (\cdot), \text{Tr} (\cdot)$, and $\text{Re} (\cdot)$, respectively. $\Vert \cdot \Vert$ and  ${\Vert \cdot \Vert}_F$ represent the $l_2$ norm and Frobenius norm, respectively.
\vspace{-5mm}

\section{System Model} \label{230104a}
\begin{figure} 
	\centering 
	\includegraphics[width=3in,height=2.1in]{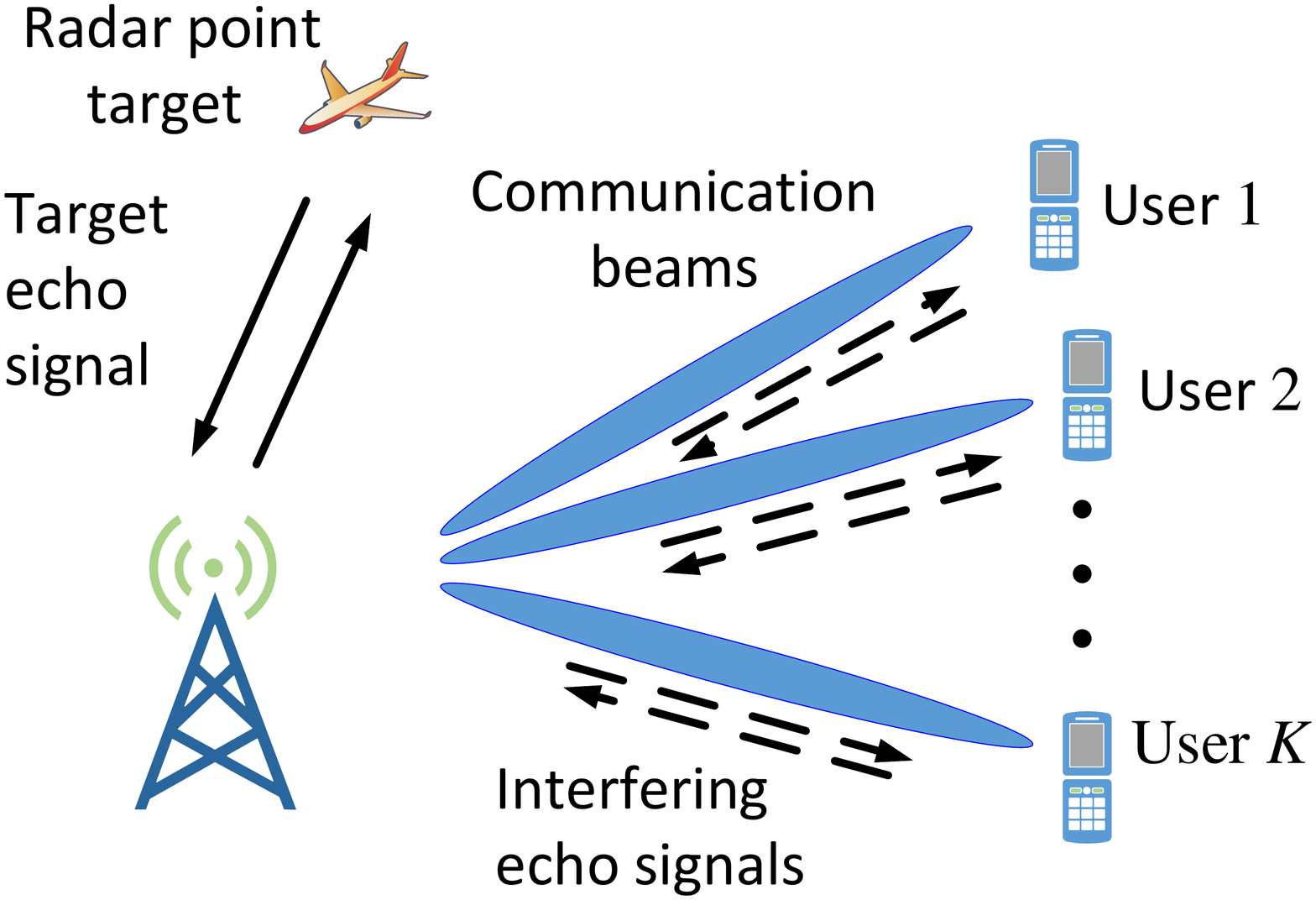}
	\caption{Integrated sensing and communication system model.}
	\label{system} 
\end{figure}
As shown in Fig. \ref{system}, we consider a DFRC system, wherein the MIMO BS equipped with $N_T$ transmit antennas and $N_R$ receive antennas communicates with $K$ single-antenna users and estimates parameters of  a target simultaneously.
Let $\mathbf{S}\in \mathbb{C}^{K \times L}$ be $K$ 
 data streams transmitting to $K$ communication users during $L$ time slots. 
We assume that each entry in $\mathbf{S}$ is i.i.d 
and follow the Gaussian distribution with zero mean and unit variance. By using the law of large numbers, when $L$ is asymptotically large, we have \cite{CRB}
\begin{align}
  \frac{1}{L} \mathbf{S} \mathbf{S}^H  \approx \mathbf{I}_K  .  \label{approx}
\end{align}	
 Let $\mathbf{W} = \left[ \mathbf{w}_1, \mathbf{w}_2, ... , \mathbf{w}_K  \right] \in \mathbb{C}^{N_T \times K}$ denote the dual-functional beamforming matrix and the transmit DFRC signal  is given by
 \begin{align}
 	\mathbf{X} = \mathbf{W} \mathbf{S}.  \label{221229a}
 \end{align} 
\vspace{-10mm}
\subsection{The Communication Model}
In terms of achieving the function of communication from the BS to the $K$ users, the received signal of the downlink communication by the users is given by
\begin{align}
	\mathbf{Y}_C 
	             & = \mathbf{H} \mathbf{X} + \mathbf{N},
\end{align} 
where $\mathbf{H} = \left[ \mathbf{h}_1, \mathbf{h}_2, ... , \mathbf{h}_K  \right]^H \in  \mathbb{C}^{K \times N_T} $  denotes the channel matrix between the BS and communication users, 
and 
$\mathbf{N} \in \mathbb{C}^{K \times L}$ denotes the additive white Gaussian noise (AWGN) matrix, i.e., $\mathbf{n} = \text{vec}(\mathbf{N}^H) \sim \mathcal{CN} (\mathbf{0},\sigma_{N}^2 \mathbf{I}_{KL})$, where $\text{vec}(\cdot)$ denotes the vectorization of a matrix which results in a column vector by stacking the matrix columns on top of one another. 
\vspace{-5mm}
\subsection{The Radar Model}
As for the function of radar, we consider the BS as a colocated MIMO radar so that the angle of departural (AoD) and the angle of arrival (AoA) of each signal are the same.
In ISAC systems, downlink communication users may inevitably become interfering scatterers in radar sensing, and the interfering echo signal generated by the communication users can affect the accuracy of radar target estimation and detection \cite{tangrelative}, \cite{tang_an}. To consider the interference problem, the DFRC BS is assumed to transmit the signal in (\ref{221229a}) and receive the reflected signals from both the target scatterer and the interfering scatterers of the communication users. 
Then, the echo signal received at the DFRC BS is given by   
\begin{align} 
	\mathbf{Y}_R &=  \mathbf{G}_R \mathbf{X} + \mathbf{G}_C \mathbf{X} + \mathbf{Z}\nonumber\\
	&=  \mathbf{G}_R \mathbf{W} \mathbf{S} +  \mathbf{G}_C \mathbf{W} \mathbf{S} + \mathbf{Z},
\end{align}
where  $\mathbf{G}_R \in \mathbb{C}^{N_R \times N_T}$ and $\mathbf{G}_C \in \mathbb{C}^{N_R \times N_T}$ denote the target response matrix and interference response matrix, respectively, and $\mathbf{Z} \in \mathbb{C}^{N_R \times L}$ denotes the AWGN, i.e., each entry of $\mathbf{Z}$ is i.i.d and follows the Gaussian distribution with zero mean and  variance $\sigma_Z^2$.
Vectorizing the echo signal matrix $\mathbf{Y}_R^H$, we can obtain that
\begin{align} 
	\mathbf{y}_R &= \text{vec}  (\mathbf{Y}_R^H) \nonumber \\
	&= \widetilde{\mathbf{S}} \widetilde{\mathbf{W}} \mathbf{g}_R +\widetilde{\mathbf{S}} \widetilde{\mathbf{W}} \mathbf{g}_C +\mathbf{z} \nonumber \\ 
	&= \widetilde{\mathbf{X}}  \mathbf{g}_R +\widetilde{\mathbf{X}}  \mathbf{g}_C +\mathbf{z},
\end{align}
where $ \widetilde{\mathbf{X}}=\widetilde{\mathbf{S}} \widetilde{\mathbf{W}}$, $\widetilde{\mathbf{S}}= \mathbf{I}_{N_R} \otimes \mathbf{S}^H$, $\widetilde{\mathbf{W}}= \mathbf{I}_{N_R} \otimes \mathbf{W}^H$,   $\mathbf{g}_R = \text{vec}(\mathbf{G}_R^H)$, $\mathbf{g}_C = \text{vec}(\mathbf{G}_C^H)$, $\mathbf{z}=\text{vec}(\mathbf{Z}^H)\sim \mathcal{CN} (\mathbf{0},\sigma_{Z}^2 \mathbf{I}_{LN_R})$.
We assume that $\mathbf{g}_R \sim \mathcal{CN} (\mathbf{0}, \mathbf{R}_R)$ and $\mathbf{g}_C \sim \mathcal{CN} (\mathbf{0}, \mathbf{R}_C)$.  Furthermore, we assume that $\mathbf{g}_R$ and $\mathbf{g}_C$ are independent to each other and also independent to $(\mathbf{z}, \mathbf{S})$. To calculate the covariance matrix of $\mathbf{g}_R$ and $\mathbf{g}_C$, we first introduce the response matrices of two types of scatterers: point scatterer and extended scatterer \cite{tangboradar}.


1) For a point scatterer, we assume that a scatterer is far away from the DFRC BS or has a small scattering surface, such as mobile phones, then it  can be regarded as a point scatterer. The corresponding response matrix can be modeled by 
\begin{align}
  \mathbf{G}_{\mathrm{point}} =\alpha \mathbf{b}(\theta) \mathbf{a}^{H}(\theta), \label{point}
\end{align}
where $\alpha$ represents the reflection coefficient, $\theta$ is the AOA/AOD of the scatterer relative to the DFRC BS, and 
$ \mathbf{a}(\theta) \in \mathbb{C}^{N_T \times 1} $ and $\mathbf{b}(\theta) \in \mathbb{C}^{N_R \times 1}$  are transmit and receive array steering vectors, respectively. 

2) For an extended scatterer, the scatterer is assumed to be close to the DFRC BS or has a big scattering surface, such as vehicles, then it can be regarded as
a surface, of which the  response matrix can be modeled by the sum of $M$ point-like scatterers, i.e., 
\begin{align}
	\mathbf{G}_{\mathrm{extended}}  =\sum_{i=1}^{M}  \alpha_i \mathbf{b}(\theta_i) \mathbf{a}^{H}(\theta_i), \label{extended}
\end{align}
where $\alpha_{i}$ and $\theta_{i}$ represent the reflection coefficient and the AoA/AoD of the $i$-th point-like scatterer,  and $ \mathbf{a}(\theta_i) \in \mathbb{C}^{N_T \times 1} $ and $\mathbf{b}(\theta_i) \in \mathbb{C}^{N_R \times 1}$ are transmit and receive array steering vectors of the $i$-th point-like scatterer, respectively.


For the rest of this paper, we assume that the sensing target is a point scatterer and investigate different types of interfering scatterer. The target response matrix of the point target $\mathbf{G}_R$ takes on the form of $\mathbf{G}_\mathrm{point}$. Thus, the covariance matirx $\mathbf{R}_R$ of $\mathbf{g}_{R}=\text{vec}(\mathbf{G}_R^H)$ is given by
\begin{align}
	\mathbf{R}_R &= \mathbb{E} (\mathbf{g}_R \mathbf{g}_R^H) \nonumber \\
	&= \beta^2 (\mathbf{b}^*(\theta) \otimes \mathbf{a}(\theta)) (\mathbf{b}^*(\theta) \otimes \mathbf{a}(\theta))^H,  \label{pointcov}
\end{align}
where $\theta$ is the AOA/AOD of the point target scatterer and
$\beta^2 = \mathbb{E} (\alpha \alpha^*) $ denotes the average strength of 
the point target echo signal. 



For the case where $K=1$, i.e., there is only one communication user, the interference response matrix $\mathbf{G}_C$ can be modeled according to the distance between the user and the DFRC BS. When the user is far away from the DFRC BS, it is treated as a point scatterer as \eqref{point}. Hence, the interference response matrix and the corresponding covariance matrix are respectively given by
\begin{align}
		\mathbf{G}_C & = \kappa \mathbf{b}(\theta^c) \mathbf{a}^{H}(\theta^c), \label{point_C}  \\   
		\mathbf{R}_C & = \mathbb{E} (\mathbf{g}_C \mathbf{g}_C^H) = \gamma^2 (\mathbf{b}^*(\theta^c) \otimes \mathbf{a}(\theta^c)) (\mathbf{b}^*(\theta^c) \otimes \mathbf{a}(\theta^c))^H  \label{pointcov_C} 
\end{align}
where  $\kappa$ and $\theta^{c}$ denote  the reflection coefficient and the AoA/AoD of the interfering echo signal, and $ \gamma^2 = \mathbb{E} (\kappa \kappa^*)  $ denotes the average strength of the point interfering echo signal. 
On the other hand, when the user is close to the DFRC BS, it is treated as an extended scatterer. Following \eqref{extended}, the interference response matrix and the corresponding covariance matrix are respectively given by
\begin{align}
	\mathbf{G}_C &=\sum_{i=1}^{M}  \kappa_i \mathbf{b}(\theta_i^c) \mathbf{a}^{H}(\theta_i^c), \label{extended_C}  \\
	\mathbf{R}_C &=  \mathbb{E} (\mathbf{g}_C \mathbf{g}_C^H) = \sum_{i=1}^{M} \gamma_i^2 (\mathbf{b}^*(\theta_i^c) \otimes \mathbf{a}(\theta_i^c)) (\mathbf{b}^*(\theta_i^c) \otimes \mathbf{a}(\theta_i^c))^H,  \label{extendedcov_C} 
\end{align}
where $\kappa_{i}$ and $\theta_{i}^{c}$ represent the reflection coefficient and the AoA/AoD of the $i$-th point-like interfering scatterer.

For the case of $K>1$, i.e., there are more than one communication user in the system,
according to the distance between the BS and the users, every user can be regarded as the point or extended scatterer. 
Then, the radar interferences caused by all of communication users can be modeled as the extended scatterer \eqref{extended_C}.

In the considered system, we aim to design the DFRC beamforming matrix $\mathbf{W}$ to meet the requirements of S\&C at the same time, wherein their different performance metrics are discussed in the following subsections. 
\vspace{-5mm}
\subsection{Communication Performance Metric}
To evaluate and guarantee the quality of service (QoS) for communication users, we utilize the transmission rate as the performance metric. Then, the QoS requirement of the $k$-th user can be represented as
\begin{align}
	R_k 
	= \log_2 \Big(1+ \frac{{\lvert \mathbf{h}_k^H \mathbf{w}_k \rvert}^2}{\sum_{j=1,j \neq k}^{K}{\lvert \mathbf{h}_k^H \mathbf{w}_j \rvert}^2 + \sigma_{N}^2} \Big) \geq r_k,  \label{rate}
\end{align}
where $R_k$ is the achievable transmission rate of the $k$-th user, and $r_k$ is the target rate required by the $k$-th user.
\vspace{-5mm}

\subsection{Radar Performance Metric}
For the performance metric of radar sensing, we choose the MI between the echo signal of the DFRC BS, $\mathbf{y}_R$, and the target response matrix, $\mathbf{g}_R$.  According to \cite[Propositions 1 and 2]{mi}, we can maximize the MI between the parameters of interest and the measurements, so as to  obtain more information about the  measured  object and reduce the measurement error, which finally results in more accurate radar estimation, detection, and classification. In DFRC systems, the BS knows the transmit signal $\mathbf{X}$ so that the MI between $\mathbf{y}_R$ and $\mathbf{g}_R$ given $\mathbf{X}$ can be written as \cite{cover}
\begin{align}
	&\quad \mathcal{I} (\mathbf{y}_R ; \mathbf{g}_R \vert \mathbf{X} = \mathbf{W} \mathbf{S} ) \nonumber \\
	&= \mathcal{H} (\mathbf{y}_R \vert \mathbf{X} = \mathbf{W} \mathbf{S}) -
	\mathcal{H} (\mathbf{y}_R \vert  \mathbf{g}_R, \mathbf{X} = \mathbf{W} \mathbf{S}) \nonumber \\
    &= - \int p(\mathbf{y}_R \vert \mathbf{X} = \mathbf{W} \mathbf{S})	\log \; p(\mathbf{y}_R \vert \mathbf{X} = \mathbf{W} \mathbf{S})   + \int p(\mathbf{y}_R \vert \mathbf{g}_R, \mathbf{X} = \mathbf{W} \mathbf{S})  \log \; p(\mathbf{y}_R \vert \mathbf{g}_R, \mathbf{X} = \mathbf{W} \mathbf{S}) , \\
	& = \log \left[  \det(\widetilde{\mathbf{S}} \widetilde{\mathbf{W}} (\mathbf{R}_R + \mathbf{R}_C ) \widetilde{\mathbf{W}}^H  \widetilde{\mathbf{S}}^H  + \sigma_{Z}^2 \mathbf{I}_{LN_R}) \right] - \log \left[  \det( \widetilde{\mathbf{S}} \widetilde{\mathbf{W}} \mathbf{R}_C \widetilde{\mathbf{W}}^H  \widetilde{\mathbf{S}}^H + \sigma_{Z}^2 \mathbf{I}_{LN_R})   \right] \nonumber \\
    & = \log \left[  \xi \det( \widetilde{\mathbf{W}} (\mathbf{R}_R + \mathbf{R}_C ) \widetilde{\mathbf{W}}^H  \widetilde{\mathbf{S}}^H \widetilde{\mathbf{S}} + \sigma_{Z}^2 \mathbf{I}_{KN_R}) \right] - \log \left[ \xi  \det( \widetilde{\mathbf{W}} \mathbf{R}_C \widetilde{\mathbf{W}}^H  \widetilde{\mathbf{S}}^H \widetilde{\mathbf{S}} + \sigma_{Z}^2 \mathbf{I}_{KN_R})   \right] \label{221230d}\\
    & = \log \left[ \xi \det( \widetilde{\mathbf{W}} (\mathbf{R}_R + \mathbf{R}_C ) \widetilde{\mathbf{W}}^H  (\mathbf{I}_{N_R} \otimes \mathbf{S})(\mathbf{I}_{N_R} \otimes \mathbf{S}^H) + \sigma_{Z}^2 \mathbf{I}_{KN_R}) \right]  \nonumber \\
    & \quad - \log \left[ \xi \det( \widetilde{\mathbf{W}} \mathbf{R}_C \widetilde{\mathbf{W}}^H  (\mathbf{I}_{N_R} \otimes \mathbf{S})(\mathbf{I}_{N_R} \otimes \mathbf{S}^H) + \sigma_{Z}^2 \mathbf{I}_{KN_R})   \right] \nonumber \\
    & \approx \log \left[  \det(  L \widetilde{\mathbf{W}} (\mathbf{R}_R + \mathbf{R}_C ) \widetilde{\mathbf{W}}^H   + \sigma_{Z}^2 \mathbf{I}_{KN_R}) \right]  - \log \left[  \det( L \widetilde{\mathbf{W}} \mathbf{R}_C \widetilde{\mathbf{W}}^H   + \sigma_{Z}^2 \mathbf{I}_{KN_R})   \right] \label{MI}. 
\end{align}
where $\xi= (\sigma_{Z}^2)^{LN_R-KN_R}$, $\mathcal{H}(\cdot)$ represents the differential entropy, 
$p(\cdot)$ represents the probability density function (PDF), (\ref{221230d}) follows from the property of the matrix determinant $\det (\mathbf{I}_m+\mathbf{AB}) = \det (\mathbf{I}_n+\mathbf{BA})$ \cite{zxd}, (\ref{MI}) follows from the property of the Kronecker product $ (\mathbf{AB}) \otimes (\mathbf{CD}) = (\mathbf{A} \otimes \mathbf{C})(\mathbf{B} \otimes \mathbf{D}) $ \cite{zxd}, and \eqref{approx} under the assumption of sufficiently large $L$.

 The aim of the paper is to design the ISAC beamforming matrix that maximizes the MI while satisfying the QoS requirements of the communication users. We will consider two cases: one is $K=1$, i.e., there is a single communication user, and the other case is $K>1$. In the case of $K=1$, since there is only one user, for notational simplicity, we will use $R$ to denote its achievable transmission rate and $r$ to denote the target rate required by the communication user, as in (\ref{rate}). Furthermore, the beamforming matrix $\mathbf{W}$ degenerates to the beamforming column vector $\bf w$  for $K=1$.
 \vspace{-5mm} 
%
%

\section{Single Communication User: beamforming design without radar interference } \label{221230}

In this section, we assume that there is a single communication user, i.e., $K=1$, and furthermore, 
the average strength of the interfering echo signal from the communication user is very weak and thus, ${\bf R}_{C}$ in \eqref{MI} can be ignored. We would like to maximize the MI of the radar target under the QoS constraint of the communication user, i.e., the formulated problem is given by
\begin{subequations} \label{1}
	\begin{align}
		\max_{\mathbf{w}} \quad  &\log \left[ \xi \det(  L \widetilde{\mathbf{W}} \mathbf{R}_R \widetilde{\mathbf{W}}^H   + \sigma_{Z}^2 \mathbf{I}_{N_R}) \right] \label{1a} \\
		{\rm s.t.} \quad & {\Vert \mathbf{w} \Vert}^2 \leq P_0 \label{1b} \\
		&  	R \geq r. \label{1c}
	\end{align}
\end{subequations}
Here, \eqref{1b} is the power constraint for the BS with $P_{0}$ being the maximum transmit power, and (\ref{1c}) follows from (\ref{rate}). 

First, 
the constraint \eqref{1c} can be simplified to
\begin{align}
	{\lvert \mathbf{h}^H \mathbf{w} \rvert}^2 \geq \Omega, \label{single SINR}
\end{align}
where $ \Omega = (2^r -1) \sigma_{N}^2 $.

Then, by taking (\ref{pointcov}) into (\ref{1a}), we have the following transformation on the objective function
\begin{align}
	&\quad\log \left[ \xi \det(  L  \widetilde{\mathbf{W}} \mathbf{R}_R \widetilde{\mathbf{W}}^H   + \sigma_{Z}^2 \mathbf{I}_{N_R}) \right] 
	 = \log\left[ (\sigma_{Z}^2)^{LN_R - 1}  (  L \beta^2  \mathbf{w}^H \mathbf{P} \mathbf{P}^H \mathbf{w}    + \sigma_{Z}^2   )  \right], \label{1-1} 
\end{align}
where $\mathbf{P} = \mathbf{a}(\theta) \mathbf{b}^H(\theta) $,
 and
the derivation of \eqref{1-1} follows the fact that $\mathbf{b} \otimes \mathbf{a} =\text{vec} (\mathbf{a} \mathbf{b}^T)$, $\text{vec}( \mathbf{A} \mathbf{B} \mathbf{C}) = (\mathbf{C}^T \otimes \mathbf{A}) \text{vec}(\mathbf{B})$, 
$ \det(\mathbf{A}) = \det(\mathbf{A}^*)$ for a Hermitian matrix, and the property of the matrix determinant $\det (\mathbf{I}_m+\mathbf{AB}) = \det (\mathbf{I}_n+\mathbf{BA})$ \cite{zxd}.
Further substituting $\mathbf{P} = \mathbf{a}(\theta) \mathbf{b}^H(\theta)$ into \eqref{1-1}, we have
\begin{align}
	\log \left[(\sigma_{Z}^2)^{LN_R - 1} (   L \beta^2 N_R \mathbf{w}^H  \mathbf{a}(\theta)  \mathbf{a}^H(\theta)  \mathbf{w}    + \sigma_{Z}^2   )\right], \label{1-ex}
\end{align}
where $  \mathbf{b}^H(\theta)  \mathbf{b}(\theta) = N_R$.
Then, ignoring the constants in \eqref{1-ex}, we only need to maximize $\mathbf{w}^H  \mathbf{a}(\theta)  \mathbf{a}^H(\theta)  \mathbf{w} $.

Therefore, problem \eqref{1} can be rewritten as follows
\begin{subequations} \label{1-2}
	\begin{align}
		\max_{\mathbf{w}} \quad  &  \mathbf{w}^H \mathbf{a}(\theta)  \mathbf{a}^H(\theta) \mathbf{w} \label{1-2a} \\
		\text{s.t.} \quad & \eqref{1b}, \thinspace \eqref{single SINR}.
	\end{align}
\end{subequations}
Problem (\ref{1-2}) is a non-convex problem, and a suboptimal solution may be found by using semidefinite relaxation (SDR) or successive convex approximation (SCA) based on CVX. However, we provide a closed-form suboptimal solution with much less complexity in Lemma \ref{lemma-1}. As shown in Appendix \ref{proof1} and Lemma 2 in \cite{proof1}, the solution in Lemma \ref{lemma-1} is obtained using the Lagrange multiplier method which can find a local optimal value of a function subject to equality and inequality constraints. From \cite[Proposition 1.1.2]{nonlinear}, the local optimal point of a convex optimization problem is also the global optimal point.
However, Problem \eqref{1-2} is a  nonconvex optimization problem, so the closed-form solution given in Lemma \ref{lemma-1} is suboptimal.\vspace{-3mm}
\begin{lemma}  \label{lemma-1}
	A suboptimal solution to problem \eqref{1-2} is given by
	\begin{align*}
		\begin{split}
			\mathbf{w} = \left \{
			\begin{array}{lr}
				\sqrt{P_0} \frac{\mathbf{a}(\theta) }{\Vert \mathbf{a}(\theta)  \Vert},  & {\lvert \mathbf{h}^H \mathbf{a}(\theta) \rvert}^2 > \frac{ \Omega {\Vert \mathbf{a}(\theta)  \Vert}^2 }{P_0} \\
				\sqrt{P_0} \frac{\mathbf{h} }{\Vert \mathbf{h}  \Vert},     &  \frac{\lvert \mathbf{a}^H(\theta) \mathbf{h} \rvert}{\Vert \mathbf{a}(\theta)  \Vert \Vert \mathbf{h}  \Vert } =  1 \\
				z_1 \frac{\mathbf{h}}{\Vert \mathbf{h} \Vert} + z_2 \frac{\mathbf{a}(\theta)}{\Vert \mathbf{a}(\theta) \Vert},                                 & \text{otherwise},
			\end{array}
			\right.
		\end{split}
	\end{align*}
	where $z_1 = \sqrt{P_0} ( \sqrt{t} - u_2 r ) \frac{\mathbf{a}^T \mathbf{h}^*}{| \mathbf{a}^H \mathbf{h} | }, z_2 = \sqrt{\frac{P_0 (1-t) }{1-r^2}}, t =  \frac{\Omega}{P_0 {\Vert \mathbf{h}  \Vert}^2}, r = \frac{{\lvert \mathbf{a}^H(\theta) \mathbf{h} \rvert} }{  {\Vert \mathbf{a}(\theta)  \Vert} {\Vert \mathbf{h}  \Vert} }$,  and $u_2 = \sqrt{\frac{1-t}{1-r^2}}$. \label{lemma1}
\end{lemma}
\textbf{\textit{Proof: }} Please refer to Appendix \ref{proof1}.  \hspace{10cm} $\blacksquare$

\section{Single Communication User: beamforming design with a point interference } \label{point-inter}
When the scattered signal reflected by the communication user can be received by the DFRC BS and its interference to radar sensing cannot be ignored, the design of the ISAC beamforming not only needs to balance the tradeoff between radar sensing and communication performance through beam resource allocation, but also needs to suppress the interference generated by the communication user on radar target sensing. In this case, we solve the following optimization problem
\begin{subequations} \label{2}
	\begin{align}
		\max_{\mathbf{w}} \quad  &\log \left[  \det(  L \widetilde{\mathbf{W}} (\mathbf{R}_R + \mathbf{R}_C ) \widetilde{\mathbf{W}}^H   + \sigma_{Z}^2 \mathbf{I}_{KN_R}) \right]   - \log \left[  \det( L \widetilde{\mathbf{W}} \mathbf{R}_C \widetilde{\mathbf{W}}^H   + \sigma_{Z}^2 \mathbf{I}_{KN_R})   \right] \label{2a} \\
		{\rm s.t.} \quad & {\Vert \mathbf{w} \Vert}^2 \leq P_0 \label{2b} \\
		&  	R \geq r. \label{2c}
	\end{align}
\end{subequations}

In this section, the communication user is assumed to be far away from the BS 
or has a small area, so that the interfering echo signal can be regarded as a point scatterer, as shown in \eqref{point_C} and \eqref{pointcov_C}. Hence, the covariance matrix of the interfering point scatterer $\mathbf{R}_C$ in Problem (\ref{2})  is given by \eqref{pointcov_C}.

Different from Problem  \eqref{1}, Problem \eqref{2} is difficult to tackle due to the existence of  the interfering echo signal in the objective function. In the following, we transform the nonconvex problem of \eqref{2} into an SDP problem by using a series of mathematical transformations.
First, the determinant of a matrix is a high-degree polynomial, which has high complexity, thus we transform the determinants in \eqref{2a} into scalars.
Motivated by \eqref{1-1}, we use the same mathematical properties to tackle \eqref{2a}, as follows
\begin{subequations} \label{2-1}
	\begin{align}
		&\quad \log \left[  \det(  L \widetilde{\mathbf{W}} (\mathbf{R}_R + \mathbf{R}_C ) \widetilde{\mathbf{W}}^H   + \sigma_{Z}^2 \mathbf{I}_{KN_R}) \right] 
		 - \log \left[  \det( L \widetilde{\mathbf{W}} \mathbf{R}_C 
		\widetilde{\mathbf{W}}^H   + \sigma_{Z}^2 \mathbf{I}_{KN_R})   \right] \nonumber \\
		& = \log \left[  \det (  L \beta^2 \mathbf{P}^H \mathbf{w} \mathbf{w}^H \mathbf{P}  + L \gamma^2 \mathbf{Q}^H \mathbf{w} \mathbf{w}^H \mathbf{Q} + \sigma_{Z}^2 \mathbf{I}_{N_R} ) \right] 	- \log \left[  \det (  L \gamma^2 \mathbf{Q}^H \mathbf{w} \mathbf{w}^H \mathbf{Q} + \sigma_{Z}^2 \mathbf{I}_{N_R} ) \right] \label{2-1a}\\
		& = \log \left[  \det (  L \beta^2 \mathbf{P}^H \mathbf{w} \mathbf{w}^H \mathbf{P}  + L \gamma^2 \mathbf{Q}^H \mathbf{w} \mathbf{w}^H \mathbf{Q} + \sigma_{Z}^2 \mathbf{I}_{N_R} ) \right] 
          - \log\left[ \varphi_1  (   L \gamma^2 \mathbf{w}^H \mathbf{Q} \mathbf{Q}^H \mathbf{w}  + \sigma_{Z}^2   ) \right], \label{2-1b}
	\end{align}
\end{subequations}
where $\varphi_1=(\sigma_{Z}^2)^{N_R-1}$ and $\mathbf{Q} = \mathbf{a}(\theta^c) \mathbf{b}^H(\theta^c)$.
%
%
The first term in \eqref{2-1b} cannot be simplified by directly using the property of the matrix determinant, 
hence, we need to seek some additional mathematical transformations. Specifically,  by setting parameters $\mathbf{T} = \text{Blkdiag}(\left[ \mathbf{w}, \mathbf{w}   \right])$ and  $\mathbf{M} = \left[ \sqrt{L} \gamma \mathbf{Q}^H, \sqrt{L} \beta \mathbf{P}^H  \right]^H$, 
the first term in \eqref{2-1b} can be recast as follows
\begin{subequations} \label{2-2}
	\begin{align}
		&\quad \log   \det (  L \beta^2 \mathbf{P}^H \mathbf{w} \mathbf{w}^H \mathbf{P}  + L \gamma^2 \mathbf{Q}^H \mathbf{w} \mathbf{w}^H \mathbf{Q} + \sigma_{Z}^2 \mathbf{I}_{N_R} )  \nonumber \\
		& = \log \left[ \varphi_2 \det( \mathbf{T}^H  \mathbf{M} \mathbf{M}^H \mathbf{T}  + \sigma_{Z}^2 \mathbf{I}_{2}    )\right]   	\label{2-2b} \\
		& = \log \left[\varphi_2 \big( ( L \gamma^2 \mathbf{w}^H \mathbf{Q} \mathbf{Q}^H \mathbf{w} + \sigma_{Z}^2 ) ( L \beta^2 \mathbf{w}^H \mathbf{P} \mathbf{P}^H \mathbf{w} + \sigma_{Z}^2 )  - L^2 \beta^2 \gamma^2 ( \mathbf{w}^H \mathbf{Q} \mathbf{P}^H \mathbf{w}  )   ( \mathbf{w}^H \mathbf{P} \mathbf{Q}^H \mathbf{w} ) \big)   \right], \label{2-2d}
	\end{align}
\end{subequations}
where $\varphi_2 = (\sigma_{Z}^2)^{N_R-2}$, and we use $\det (\mathbf{I}_m+\mathbf{AB}) = \det (\mathbf{I}_n+\mathbf{BA})$ in \eqref{2-2b} to transform the
determinant of an $N_R \times N_R$ matrix into the determinant of a $2 \times 2$ matrix.

Combining  \eqref{2-1b} and \eqref{2-2d}, the objective function \eqref{2a} is equivalent to
\begin{align}
	&  \log \Big[ ( L \gamma^2 \text{Tr}( \mathbf{Q} \mathbf{Q}^H \mathbf{w} \mathbf{w}^H) + \sigma_{Z}^2 ) ( L \beta^2 \text{Tr}( \mathbf{P} \mathbf{P}^H \mathbf{w} \mathbf{w}^H) + \sigma_{Z}^2 )   \nonumber \\
	&  - L^2 \beta^2 \gamma^2 \text{Tr}(  \mathbf{Q} \mathbf{P}^H \mathbf{w} \mathbf{w}^H )   \text{Tr}(  \mathbf{P} \mathbf{Q}^H \mathbf{w} \mathbf{w}^H )   \Big]  
      - \log  (   L \gamma^2 \text{Tr}(\mathbf{Q} \mathbf{Q}^H \mathbf{w} \mathbf{w}^H ) + \sigma_{Z}^2   ) - \log(\sigma_{Z}^2).  \label{2-3}
\end{align}
%
Furthermore, maximizing \eqref{2-3} is equivalent to maximizing the following expression
\begin{align}
		\frac{( L \gamma^2 \text{Tr}( \mathbf{Q} \mathbf{Q}^H \mathbf{w} \mathbf{w}^H) + \sigma_{Z}^2 ) ( L \beta^2 \text{Tr}( \mathbf{P} \mathbf{P}^H \mathbf{w} \mathbf{w}^H) + \sigma_{Z}^2 ) -   L^2 \beta^2 \gamma^2 \text{Tr}(  \mathbf{Q} \mathbf{P}^H \mathbf{w} \mathbf{w}^H )   \text{Tr}(  \mathbf{P} \mathbf{Q}^H \mathbf{w} \mathbf{w}^H )}{L \gamma^2 \text{Tr}(\mathbf{Q} \mathbf{Q}^H \mathbf{w} \mathbf{w}^H ) + \sigma_{Z}^2 }. \label{2-4}
\end{align}

The problem of maximizing \eqref{2-4} is still difficult to address, so 
by means of the Schur complement and an auxiliary variable, maximizing \eqref{2-4} can be reformulated as the following SDP problem
\begin{subequations} \label{2-5}
	\begin{align}
		\min_{\overline{\mathbf{W}}, t} \quad  & -t \label{2-5a} \\
		{\rm s.t.} \quad & \left[ \begin{array}{cc}
			L \beta^2 \text{Tr}( \mathbf{P} \mathbf{P}^H \overline{\mathbf{W}} ) + \sigma_{Z}^2 - t  &
			L \gamma \beta \text{Tr} (\mathbf{Q} \mathbf{P}^H \overline{\mathbf{W}}  ) \\
			L \beta \gamma \text{Tr}( \mathbf{P} \mathbf{Q}^H \overline{\mathbf{W}} ) & 
			L \gamma^2 \text{Tr}(\mathbf{Q} \mathbf{Q}^H \overline{\mathbf{W}} ) + \sigma_{Z}^2
		\end{array}  \right]  \succeq 0 \label{2-5b} \\
	& \overline{\mathbf{W}} \succeq \mathbf{0}, \text{rank}(\overline{\mathbf{W}}) = 1.  \label{2-5c}
	\end{align}
\end{subequations}
where $t$ is an auxiliary variable and $\overline{\mathbf{W}} = \mathbf{w} \mathbf{w}^H$ is a rank-one matrix.

Correspondingly, constraints \eqref{2b} and \eqref{2c} are transformed into the constraints of variable $\overline{\mathbf{W}}$, as follows
\begin{align}
	&\eqref{2b} \Rightarrow \text{Tr}( \overline{\mathbf{W}} ) \leq P_0 , \label{2-6} \\
	&\eqref{2c} \Rightarrow \eqref{single SINR} \Rightarrow \text{Tr}( \mathbf{h} \mathbf{h}^H \overline{\mathbf{W}} ) \geq \Omega. \label{2-7}
\end{align}

Finally, with \eqref{2-5}-\eqref{2-7}, the optimization problem in \eqref{2} can be recast as follows
\begin{subequations} \label{2-8}
	\begin{align}
		\min_{\overline{\mathbf{W}}, t} \quad  & -t \label{2-8a} \\
       {\rm s.t.} \quad & \eqref{2-5b}, \thinspace \eqref{2-5c}, \thickspace \eqref{2-6}, \thinspace \eqref{2-7}. \label{2-8b}
	\end{align}
\end{subequations}

By applying the SDR technique to drop the nonconvex constraint $\text{rank}(\overline{\mathbf{W}}) = 1$, the problem in \eqref{2-8} can be relaxed into a convex SDP problem and solved using CVX. 
Finally, the Gaussian randomization technique can be used to obtain the feasible rank-one solution of Problem \eqref{2-8}, which is suboptimal \cite{sdr}. The computational complexity of Problem \eqref{2-8} is approximately given by $\mathcal{O} ( \text{max}{\{ N_T, 3 \}}^4 N_T^{\frac{1}{2}}\log(\frac{1}{\zeta} )   )$, where  $\zeta$ is the given solution accuracy \cite{sdr}.
\vspace{-5mm}

\section{Single Communication User: beamforming design with an extended interference} \label{221230b}
In this section, we investigate 
the case where 
the single communication user is treated
as an extended scatterer 
such as a vehicle, in radar sensing.
The resulting optimization problem is still given by \eqref{2} in Section \uppercase\expandafter{\romannumeral4}, 
except that the covariance matrix of the interference response matrix, $\mathbf{R}_C$, is given by \eqref{extendedcov_C}. The high-rank matrix, $\mathbf{R}_C$, increases the challenge to tackle the objective function in \eqref{2a}.
 Thus, we resort to the Majorization-Minimization (MM) method \cite{mm}. 
 \vspace{-5mm}
\subsection{Majorization-Minimization Method}
MM is an efficient technique to tackle the optimization problems with complex nonconvex objective function and simple constraints. 
The purpose of MM is to construct an easy-to-solve upper bound (lower bound) surrogate function of the original objective function for a minimization (maximization) problem.
For instance,  in a maximization problem, we need to find a minorizer, $h(\mathbf{x}| \mathbf{x}^{(\upsilon)}) $, of the original nonconcave objective function, $g(\mathbf{x})$, for a given point  $ \mathbf{x}^{(\upsilon)} $. In order to ensure convergence,  $h(\mathbf{x}| \mathbf{x}^{(\upsilon)}) $ must satisfy the following conditions:
\begin{align*}
\mathrm{(A1):} \thinspace & h(\mathbf{x}^{(\upsilon)} | \mathbf{x}^{(\upsilon)}) = g (\mathbf{x}^{(\upsilon)}) , \quad \forall \mathbf{x}^{(\upsilon)} \in \mathcal{X}; \\
\mathrm{(A2):} \thinspace& h(\mathbf{x} | \mathbf{x}^{(\upsilon)}) \leq g (\mathbf{x}) , \quad \forall \mathbf{x},\mathbf{x}^{(\upsilon)} \in \mathcal{X}; \\
\mathrm{(A3):} \thinspace& h^{'}(\mathbf{x} | \mathbf{x}^{(\upsilon)}; \mathbf{d})|_{\mathbf{x}=\mathbf{x}^{(\upsilon)}} = g^{'} (\mathbf{x}^{(\upsilon)};\mathbf{d}), \quad \forall \mathbf{d} \thinspace\thinspace \text{with} \thinspace\thinspace \mathbf{x}^{(\upsilon)} + \mathbf{d} \in \mathcal{X}; \\
\mathrm{(A4):} \thinspace & h(\mathbf{x} | \mathbf{x}^{(\upsilon)}) \thinspace\thinspace \text{is continuous in} \thinspace\thinspace \mathbf{x} \thinspace\thinspace \text{and}  \thinspace\thinspace \mathbf{x}^{(\upsilon)},
\end{align*}
where $\mathcal{X}$ is a convex set and $g^{'} (\mathbf{x}^{(\upsilon)};\mathbf{d})$ is direction derivative of $g(\mathbf{x}^{(\upsilon)})$ in the direction of $\mathbf{d}$. 
 \vspace{-5mm}
\subsection{Beamforming Design Based on the MM Method}
Before using the MM method, we first explore the inherent concavity and convexity of \eqref{2a} via some transformations. Specifically, let $g(\bf w) $ be the objective function in \eqref{2a}, which is first rewritten as follows
\begin{subequations} \label{3-1}
\begin{align}
	  g(\mathbf{w}) 
	 &=  \log   \det(  \delta \widetilde{\mathbf{W}} \mathbf{R}_{RC} \widetilde{\mathbf{W}}^H   +  \mathbf{I}_{KN_R})  - \log   \det( \delta \widetilde{\mathbf{W}} \mathbf{R}_C \widetilde{\mathbf{W}}^H   +  \mathbf{I}_{KN_R})   \label{3-1b} \\
	& = - \log   \det \left( \delta  \mathbf{R}_{R}^{\frac{1}{2},H} \widetilde{\mathbf{W}}^H (\delta\widetilde{\mathbf{W}}\mathbf{R}_{RC} \widetilde{\mathbf{W}}^H+\mathbf{I}_{KN_R})^{-1} \widetilde{\mathbf{W}}\mathbf{R}_R^{\frac{1}{2}} + \mathbf{I}_{N_T N_R}   \right)^{-1}   ,  \label{3-1c}
\end{align}
\end{subequations}
where $\delta = \frac{L}{\sigma_{Z}^2}$ and $\mathbf{R}_{RC} = \mathbf{R}_R + \mathbf{R}_C$, and (\ref{3-1c}) follows from the properties of the determinant $\det(\mathbf{A}^{-1}) = \frac{1}{\det(\mathbf{A})}$ and $ \det(\mathbf{A}\mathbf{B}) = \det(\mathbf{A})\det(\mathbf{B})$ \cite{zxd}.

By using the Woodbury matrix identity \cite{zxd}, i.e., $ (\mathbf{A}+\mathbf{U}\mathbf{C}\mathbf{V})^{-1} = \mathbf{A}^{-1} - \mathbf{A}^{-1}\mathbf{U}(\mathbf{C}^{-1}+\mathbf{V}\mathbf{A}^{-1}\mathbf{U})^{-1}\mathbf{V}\mathbf{A}^{-1} $, we can rewrite \eqref{3-1c} as follows 
\begin{align}
 g(\widetilde{\mathbf{W}},\mathbf{T}) =	- \log \det ( \mathbf{I}_{N_TN_R} - \delta \mathbf{R}_{R}^{\frac{1}{2},H} \widetilde{\mathbf{W}}^H \mathbf{T}^{-1}  \widetilde{\mathbf{W}}\mathbf{R}_R^{\frac{1}{2}}  ), \label{3-2}
\end{align} 
where $\mathbf{T} = \mathbf{I}_{KN_R} + \delta\widetilde{\mathbf{W}}\mathbf{R}_{RC} \widetilde{\mathbf{W}}^H  $.
Since $\mathbf{R}_R \succeq 0$ and $\mathbf{R}_C \succeq 0$, we have $\mathbf{R}_{RC} \succeq 0$.
Thus, following from the property of a positive definite matrix, we can obtain $\mathbf{T} \succ 0$ and $\mathbf{R}_R^{\frac{1}{2}} \succeq 0$. According to Lemma 3 in \cite{palomar}, $g(\widetilde{\mathbf{W}},\mathbf{T})$ given in  \eqref{3-2} is jointly convex over matrices $(\widetilde{\mathbf{W}},\mathbf{T})$ when $\mathbf{R}_R^{\frac{1}{2}}$ is a positive semidefinite matrix.
Further using the first-order Taylor expansion,  the lower bound $ h (\widetilde{\mathbf{W}}, \mathbf{T} | \widetilde{\mathbf{W}}^{(\upsilon)},  \mathbf{T}^{(\upsilon)} )$ of $g(\widetilde{\mathbf{W}},\mathbf{T})$ at given points $\widetilde{\mathbf{W}}^{(\upsilon)},\mathbf{T}^{(\upsilon)}$ is given by 
	\begin{align}
		g (\widetilde{\mathbf{W}}, \mathbf{T} ) &\geq  h (\widetilde{\mathbf{W}}, \mathbf{T} | \widetilde{\mathbf{W}}^{(\upsilon)},  \mathbf{T}^{(\upsilon)} ) \nonumber \\
		&= g (\widetilde{\mathbf{W}}^{(\upsilon)}, \mathbf{T}^{(\upsilon)} )  + \text{Tr} \left(  \left( \frac{\partial g }{\partial \widetilde{\mathbf{W}}} \vert_{\widetilde{\mathbf{W}}^{(\upsilon)}}  \right)^H \left( \widetilde{\mathbf{W}} - \widetilde{\mathbf{W}}^{(\upsilon)}  \right)  \right)	\nonumber \\
		& \quad +  \text{Tr} \left(  \left( \frac{\partial g }{\partial \widetilde{\mathbf{W}}^*} \vert_{\widetilde{\mathbf{W}}^{(\upsilon),*}}  \right)^H \left( \widetilde{\mathbf{W}}^* - \widetilde{\mathbf{W}}^{(\upsilon),*}  \right)  \right) +  \text{Tr} \left(  \left( \frac{\partial g }{\partial \mathbf{T}} \vert_{\mathbf{T}^{(\upsilon)}}  \right)^H \left( \mathbf{T} - \mathbf{T}^{(\upsilon)}  \right)  \right) \nonumber \\
	    &= g (\widetilde{\mathbf{W}}^{(\upsilon)}, \mathbf{T}^{(\upsilon)} )  + \delta \text{Tr}\left( \mathbf{R}_R^{\frac{1}{2}} (\mathbf{M}^{(\upsilon)})^{-1} \mathbf{R}_{R}^{\frac{1}{2},H} \widetilde{\mathbf{W}}^{(\upsilon),H} (\mathbf{T}^{(\upsilon)})^{-1} ( \widetilde{\mathbf{W}} - \widetilde{\mathbf{W}}^{(\upsilon)}  )  \right) \nonumber \\
	    & \quad + \delta \text{Tr}\left(   (\mathbf{T}^{(\upsilon)})^{-1} \widetilde{\mathbf{W}}^{(\upsilon)}  \mathbf{R}_R^{\frac{1}{2}} (\mathbf{M}^{(\upsilon)})^{-1} \mathbf{R}_{R}^{\frac{1}{2},H} ( \widetilde{\mathbf{W}} - \widetilde{\mathbf{W}}^{(\upsilon)}  )^H  \right) \nonumber \\
	    & \quad - \delta \text{Tr}\left(  \mathbf{B}^{(\upsilon)}  (\mathbf{M}^{(\upsilon)})^{-1}  \mathbf{B}^{(\upsilon),H} ( \mathbf{T} - \mathbf{T}^{(\upsilon)} )  \right) ,       \label{3-3}
	\end{align}
where $\mathbf{M}^{(\upsilon)} = \mathbf{I}_{N_TN_R} - \delta (\mathbf{R}_{R}^{\frac{1}{2}})^H \widetilde{\mathbf{W}}^{(\upsilon),H} (\mathbf{T}^{(\upsilon)})^{-1}  \widetilde{\mathbf{W}}^{(\upsilon)}   \mathbf{R}_R^{\frac{1}{2}} $, and $ \mathbf{B}^{(\upsilon)} =(\mathbf{T}^{(\upsilon)})^{-1} \widetilde{\mathbf{W}}^{(\upsilon)} \mathbf{R}_R^{\frac{1}{2}} $.
%
Since $ \mathbf{T} = \mathbf{I}_{KN_R} + \delta\widetilde{\mathbf{W}}\mathbf{R}_{RC} \widetilde{\mathbf{W}}^H  $ and $  \mathbf{T}^{(\upsilon)} = \mathbf{I}_{KN_R} + \delta \widetilde{\mathbf{W}}^{(\upsilon)}\mathbf{R}_{RC} \widetilde{\mathbf{W}}^{(\upsilon),H} $, $h (\widetilde{\mathbf{W}}, \mathbf{T} | \widetilde{\mathbf{W}}^{(\upsilon)},  \mathbf{T}^{(\upsilon)} )$ is equivalent to
\begin{align}
		 h( \mathbf{w} | \mathbf{w}^{(\upsilon)}) = \delta \text{vec}^H (\mathbf{J}_1)  \text{vec} ( \widetilde{\mathbf{W}} ) + \delta \text{vec}^H ( \widetilde{\mathbf{W}} ) \text{vec}(\mathbf{J}_1)  -   \delta^2 \text{vec}^H ( \widetilde{\mathbf{W}} ) \mathbf{J}_2 \text{vec} ( \widetilde{\mathbf{W}} )  +    \mathrm{const} , \label{3-4}
\end{align}
where $ \mathbf{J}_1 = (\mathbf{T}^{(\upsilon)})^{-1} \widetilde{\mathbf{W}}^{(\upsilon)} \mathbf{R}_{R}^{\frac{1}{2}} (\mathbf{M}^{(\upsilon)})^{-1} \mathbf{R}_{R}^{\frac{1}{2},H} $, $ \mathbf{J}_2 = \mathbf{R}_{RC}^* \otimes \left( \mathbf{B}^{(\upsilon)}  (\mathbf{M}^{(\upsilon)})^{-1}  \mathbf{B}^{(\upsilon),H}  \right) $, and $  \mathrm{const} = g (\widetilde{\mathbf{W}}^{(\upsilon)}, \mathbf{T}^{(\upsilon)} )  - 2 \delta \text{Tr}(  (\mathbf{M}^{(\upsilon)})^{-1} \mathbf{R}_{R}^{\frac{1}{2},H} \widetilde{\mathbf{W}}^{(\upsilon),H} \mathbf{B}^{(\upsilon)}    )  + \delta^2 \text{Tr}(  \mathbf{B}^{(\upsilon)}  (\mathbf{M}^{(\upsilon)})^{-1}  \mathbf{B}^{(\upsilon),H} \widetilde{\mathbf{W}}^{(\upsilon)} \mathbf{R}_{RC}   \widetilde{\mathbf{W}}^{(\upsilon),H} ) $.

Hence, applying the MM method, $h( \mathbf{w} | \mathbf{w}^{(\upsilon)})$ in \eqref{3-4} is the lower bound of the original function $g(\mathbf{w})$ in \eqref{2a}. However, $\widetilde{\mathbf{W}}$ is not an arbitrary matrix because it is defined as $\widetilde{\mathbf{W}}= \mathbf{I}_{N_R} \otimes \mathbf{w}^H$. We wish to express $\widetilde{\mathbf{W}}$ in terms of $\mathbf{w}$ and perform the optimization over $\mathbf{w}$. Towards this end, we have the following lemma to show the linear relationship between the DFRC beamforming matrix $\mathbf{W}$ and $\widetilde{\mathbf{W}}$. Note that Lemma 2 is general, i.e., it is suitable for the case of single communication user, i.e., $K=1$ and $\mathbf{W}$ becomes $\mathbf{w}$, and also the case of multiple communication users. \vspace{-3mm}
\begin{lemma} \label{lemma-2}
    Let $\tilde{\mathbf{w}}=\text{vec}(\mathbf{W})$, then $\text{vec}(\widetilde{\mathbf{W}})=\text{vec}(\mathbf{I}_{N_{R}}\otimes\mathbf{W}^{H})=\mathbf{F}\tilde{\mathbf{w}}^{*} $, \\
    where $\mathbf{F} =\left[   
	\mathbf{C}_{1} \otimes \mathbf{I}_K,   
	\mathbf{C}_{2} \otimes \mathbf{I}_K,     
	\dots , 
	\mathbf{C}_{N_TN_R} \otimes \mathbf{I}_K  
	\right]^T \mathbf{K}_{N_TK}$, $\mathbf{K}_{N_TK} \in \mathbb{R}^{N_T K \times N_T K} $ is the commutation matrix, 
	 and $\mathbf{C}_i \in \mathbb{R}^{N_T \times N_R}, i=1,2,...,N_T N_R$, is an elementary matrix whose $( 1+ mod (i-1, N_T), \lceil    \frac{i}{N_T}   \rceil )$-th element is equal to one and others are all zero.	
\end{lemma}
\textbf{\textit{Proof: }} Please refer to Appendix \ref{proof2}. \hspace{10cm} $\blacksquare$

Note that in this section, we are interested in the case of a single communication user, i.e., $K=1$. In this case, Lemma \ref{lemma-2} shows that $\mathbf{w}=\tilde{\mathbf{w}}$ and 
$ \mathbf{K}_{N_TK} = \mathbf{I}_{N_T} $.

With the aid of Lemma \ref{lemma-2}, $h(\mathbf{w} | \mathbf{w}^{(\upsilon)})$ in \eqref{3-4} is equivalent to
\begin{align}
  h(\mathbf{w} | \mathbf{w}^{(\upsilon)}) =	2\delta \text{Re}(\mathbf{w}^H \mathbf{j}_1) - \delta^2 \mathbf{w}^H \mathbf{J}_3 \mathbf{w} + \mathrm{const}, \label{3-5}
\end{align}
where $\mathbf{j}_1 = \mathbf{F}^T \text{vec}^* ( \mathbf{J}_1 ) $ and $ \mathbf{J}_3 = \mathbf{F}^T  \mathbf{J}_2^* \mathbf{F} $.

Next, the nonconvex constraint \eqref{2c} is addressed using the SCA method as follows
\begin{align}
      \mathbf{w}^H \mathbf{h} \mathbf{h}^H \mathbf{w} \geq  2 \text{Re} \left(   \mathbf{w}^{(\upsilon),H} \mathbf{h} \mathbf{h}^H \mathbf{w}  \right)  - \mathbf{w}^{(\upsilon),H} \mathbf{h} \mathbf{h}^H \mathbf{w}^{(\upsilon)} \geq \Omega, \label{3-11}
\end{align}

Upon replacing the objective function of Problem \eqref{2} by \eqref{3-5}, discarding the constant in \eqref{3-5}, transforming the maximization problem \eqref{2} into the minimization problem,  and substituting the transmission rate constraint \eqref{2c} with \eqref{3-11}, we can obtain the following problem
\begin{subequations}\label{3-12}
\begin{align}
	\min_{\mathbf{w}} \quad  &	- 2 \text{Re}(\mathbf{w}^H \mathbf{j}_1) + \delta \mathbf{w}^H \mathbf{J}_3 \mathbf{w} \label{3-12a} \\
{\rm s.t.} \quad &  \mathbf{w}^H \mathbf{w} \leq P_0 \label{3-12b} \\
	&  -2 \text{Re} \left(   \mathbf{w}^{(\upsilon),H} \mathbf{h} \mathbf{h}^H \mathbf{w}  \right)  + \mathbf{w}^{(\upsilon),H} \mathbf{h} \mathbf{h}^H \mathbf{w}^{(\upsilon)} + \Omega \leq 0, \label{3-12c}
\end{align}
\end{subequations}
where $\delta$ is a constant so that we can divide the objective function by $\delta$.
Here, Problem \eqref{3-12} is a standard convex optimization problem which can be solved using CVX directly. Nonetheless, in order to reduce the computational complexity, we drive a suboptimal closed-form solution by means of the Lagrangian dual decomposition method \cite{boyd},\cite{pan}.
First, Problem \eqref{3-12} may be infeasible due to the fact that the set of constraints \eqref{3-12b} and \eqref{3-12c} may not intersect. To obtain feasible solutions, Problem \eqref{3-12} must satisfy the Slater's condition shown in Lemma \ref{lemma-3}. 
\vspace{-3mm}
\begin{lemma} \label{lemma-3}
Let the optimal value of the problem 	\begin{subequations} \label{3-13}
		\begin{align}
			\max_{\mathbf{w}} \quad  &  \mathbf{w}^H \mathbf{h} \mathbf{h}^H \mathbf{w}  \label{3-13a} \\
		{\rm s.t.} \quad & \mathbf{w}^H \mathbf{w} \leq P_0,  \label{3-13b}
		\end{align}
	\end{subequations}
be $\Omega_1$. If $\Omega_1 > \Omega$, where $\Omega =(2^r -1) \sigma_{N}^2 $, then Problem \eqref{3-12} satisfies the Slater's condition. Otherwise, Problem \eqref{3-12} is infeasible.

\end{lemma}
\textbf{\textit{Proof: }} Please refer to Appendix \ref{proof3}.  \hspace{10cm} $\blacksquare$

 According to Lemma \ref{lemma-3}, strong duality holds for Problem \eqref{3-12} if $\Omega_1 > \Omega$, which motivates us to  solve its dual problem. 
More specifically, the partial Lagrangian function of Problem \eqref{3-12} is given by 
\begin{align}
	\mathcal{L} (\mathbf{w}, \tau) = -2  \text{Re} (\mathbf{w}^H \mathbf{j}_1 ) +\delta \mathbf{w}^H \mathbf{J}_3 \mathbf{w}  +  \tau ( \mathbf{w}^H \mathbf{w} - P_0 ), \label{3-14}
\end{align}
where $\tau$ is the Lagrange multiplier associated with constraint \eqref{3-12b}.
Using \eqref{3-14}, we obtain the dual function of Problem \eqref{3-12} by solving the following problem:
\begin{subequations}\label{3-15}
\begin{align}
	q(\tau) = \min_{\mathbf{w}} \quad  & \mathcal{L} (\mathbf{w}, \tau)  \label{3-15a}  \\ 
{\rm s.t.} \quad & -2 \text{Re} \left(   \mathbf{w}^{(\upsilon),H} \mathbf{h} \mathbf{h}^H \mathbf{w}  \right)  + \widetilde{\Omega}  \leq 0 , \label{3-15b}
\end{align}
\end{subequations}
where $\widetilde{\Omega} = \mathbf{w}^{(\upsilon),H} \mathbf{h} \mathbf{h}^H \mathbf{w}^{(\upsilon)} + \Omega $.
Next, the dual problem can be written as follows
\begin{align}
	\max_{\tau} \quad  & q(\tau)    \quad   \mathrm{s.t.} \quad \tau \geq 0.   \label{3-16} 
\end{align}

In the following, 
Problem \eqref{3-15} is firstly solved with a given $\tau$, and then $\tau$ is found by solving Problem \eqref{3-16}.
%
With a given $\tau$, Problem \eqref{3-15} has a single convex constraint, which inspires us to introduce another Lagrange multiplier $\mu$ for constraint \eqref{3-15b}. The corresponding Lagrangian 
function of Problem \eqref{3-15} can be written as follows
\begin{align}
	\mathcal{L} (\mathbf{w}, \mu) = -2  \text{Re} (\mathbf{w}^H \mathbf{j}_1 ) +\delta \mathbf{w}^H \mathbf{J}_3 \mathbf{w} +   \tau ( \mathbf{w}^H \mathbf{w} - P_0 ) + \mu 
	\left( \widetilde{\Omega}   -2 \text{Re} \left(   \mathbf{w}^{(\upsilon),H} \mathbf{h} \mathbf{h}^H \mathbf{w}  \right) \right).\label{3-17}
\end{align}
By setting the first order derivative of $\mathcal{L} (\mathbf{w}, \mu)$ w.r.t $\mathbf{w}^*$ to zero, i.e., $ \frac{\partial \mathcal{L}}{\partial \mathbf{w}^* } = 0 $, the optimal solution $\hat{\mathbf{w}}$ is given by
\begin{align}
	\hat{\mathbf{w}}(\mu) = \left(  \delta \mathbf{J}_3 + \tau \mathbf{I}_{N_T}  \right)^{\dagger} \left(    \mu \mathbf{h} \mathbf{h}^H \mathbf{w}^{(\upsilon)} + \mathbf{j}_1 \right),\label{3-18}
\end{align}
where  $\hat{\mathbf{w}}(\mu) $ is a function of the Lagrange multiplier $ \mu $, and
$\tau$ is regarded as a known parameter  in \eqref{3-17} and \eqref{3-18}.  The optimal $\hat{\mathbf{w}}(\mu)$ must satisfy complementary slackness condition 
\begin{align}
  \mu  \left( \widetilde{\Omega}   -2 \text{Re} \left(   \mathbf{w}^{(\upsilon),H} \mathbf{h} \mathbf{h}^H \hat{\mathbf{w}}(\mu)  \right) \right) = 0 . \label{3-19}
\end{align}
Thus, $\mu$ satisfies
\begin{equation}\label{3-101}
	\mu=\begin{cases} 0, & \textrm{if}-2 \text{Re} \left(   \mathbf{w}^{(\upsilon),H} \mathbf{h} \mathbf{h}^H \hat{\mathbf{w}}(0)  \right)  + \widetilde{\Omega}  \leq 0,\\ 
		\frac{ \widetilde{\Omega} -  2 \text{Re} \left(   \mathbf{w}^{(\upsilon),H} \mathbf{h} \mathbf{h}^H  \left(  \delta \mathbf{J}_3  + \tau \mathbf{I}_{N_T}  \right)^{\dagger}  \mathbf{j}_1 \right)}{  2  \left(   \mathbf{w}^{(\upsilon),H} \mathbf{h} \mathbf{h}^H \left(\delta \mathbf{J}_3 + \tau \mathbf{I}_{N_T}  \right)^{\dagger} \mathbf{h} \mathbf{h}^H \mathbf{w}^{(\upsilon)}\right)}, & \textrm{otherwise}. \end{cases}
\end{equation}
Case 2 in \eqref{3-101} is due to the fact that there must exist a $\mu>0$ such that $-2 \text{Re} (   \mathbf{w}^{(\upsilon),H} \mathbf{h} \mathbf{h}^H \hat{\mathbf{w}}(\mu)  )  + \widetilde{\Omega}  = 0$.
%

Substituting $\mu$ given in \eqref{3-101} into \eqref{3-18}, the right-hand side of \eqref{3-18} is a function of $\tau$, which we denote $\overline{\mathbf{w}}(\tau)$. we can express the optimal solution $\overline{\mathbf{w}}(\tau)$ as a function of $\tau$ in Problem \eqref{3-15} and obtain the dual function $q(\tau)$.
The optimal $\tau$ of Problem \eqref{3-16} can also be calculated according to the 
 complementary slackness condition:
\begin{align}
	\tau ( \overline{\mathbf{w}}^H(\tau) \overline{\mathbf{w}}(\tau) - P_0 ) = 0. \label{3-23}
\end{align}
In the following, we define transmit power $\overline{\mathbf{w}}^H(\tau) \overline{\mathbf{w}}(\tau)$ as a function of $\tau$, i.e., $f(\tau) = \overline{\mathbf{w}}^H(\tau) \overline{\mathbf{w}}(\tau)$.

If $\tau = 0$, the optimal solution $\overline{\mathbf{w}}(0)$ needs to satisfy the power constraint \eqref{3-12b}, i.e., 
\begin{align}
	f(0) \leq P_0.  \label{3-24}
\end{align}
Otherwise, there must exist a $\tau > 0$ satisfying
\begin{align}
	 f(\tau)  = P_0, \label{3-25}
\end{align}
Note that deriving the closed-form expression of  $\tau$ directly from \eqref{3-25} is very difficult due to the complex expression of $f(\tau)$. Hence, we prove that $f(\tau)$ is a monotonically decreasing function in the following lemma, which helps us find the optimal $\tau > 0$ using the bisection search method. 
\vspace{-10mm}
\begin{lemma} \label{lemma-4} 
	The transmit power $f(\tau)$ is a monotonically decreasing function of $\tau$.
\end{lemma}
 \textbf{\textit{Proof: }} The proof is similar to \cite[Appendix A]{pan}, and is omitted due to space limitation.  \hspace{0.5cm}  $\blacksquare$

To summarize, the proposed algorithm solving Problem \eqref{2} for the case of a single communication user who is an extended interference is summarized in Algorithm \ref{alg1}. 
 \begin{algorithm}[!t]
 	\caption{MM-based Algorithm for Solving Problem \eqref{2} for a single communication user who is an extended interference }
 	\label{alg1}
 	\begin{algorithmic}[1]
 		 \STATE Initialize the tolerance $\epsilon_1$, the lower bound $\tau_l > 0$, the upper bound $\tau_u > 0$, the maximum ratio transmission vector $\mathbf{w}^{(0)}$, and the maximum iteration number ${\upsilon}_{max}$. Set iteration index $ \upsilon = 0 $. Calculate the initial objective function of Problem \eqref{2} as $g(\mathbf{w}^{(0)})$.
 		\STATE Calculate $\widetilde{\Omega}^{(\upsilon)} = \mathbf{w}^{(\upsilon) H} \mathbf{h} \mathbf{h}^H \mathbf{w}^{(\upsilon)} + \Omega $.
 		\IF {Case 1 in \eqref{3-101} and inequality \eqref{3-24} hold.}
 		\STATE Set $\tau = 0$ and $\mu = 0$. Update $ \mathbf{w}^{(\upsilon+1)} $ according to \eqref{3-18}.
        \ELSIF { Inequality \eqref{3-24} holds. }
        \STATE Set $ \tau = 0$. Calculate $\mu$ according to Case 2 in \eqref{3-101}. Update $ \mathbf{w}^{(\upsilon+1)} $ according to \eqref{3-18}.
        \ELSE
 		\REPEAT
 		\STATE Calculate $\tau = \frac{ \tau_l + \tau_u  }{2}$.
 		\STATE Calculate $\mu$ according to \eqref{3-101}.
 			Update $ \mathbf{w}^{(\upsilon+1)} $ according to \eqref{3-18}.
 		\STATE  If  $f(\tau) < P_0$, set $\tau_u = \tau$. Otherwise, set $\tau_l = \tau$. 
 		\UNTIL { $ | \tau_u -\tau_l | \leq \epsilon_1 $}.
 		\ENDIF
 		\STATE If $| \frac{g ( \mathbf{w}^{(\upsilon+1)} ) - g ( \mathbf{w}^{(\upsilon)} )}{g ( \mathbf{w}^{(\upsilon)} ) } | \leq \epsilon_1  $ or $\upsilon \geq \upsilon_{max}$, terminate.  Otherwise, set $\upsilon = \upsilon + 1 $ and go to step 2. 
 	\end{algorithmic}
 \end{algorithm}
\vspace{-5mm}
\subsection{Complexity Analysis}
  From Step 3 to 13, the main complexity of Algorithm \ref{alg1} per iteration is caused by the update of $\mathbf{w}^{(\upsilon+1)}$ and the calculation of $\mu$, i.e., \eqref{3-18} and \eqref{3-101}, respectively.
Then, the worst case complexity is using bisection search method to update  $\mathbf{w}^{(\upsilon+1)}$ and $\mu$, where it needs about $\log_2(\frac{\tau_u-\tau_l}{\epsilon_1} )$ iterations to converge,  where  $\tau_u$ and $\tau_l$ is the upper and lower bounds for calculation of $\tau$, respectively, and $\epsilon_1$ is the tolerance of Algorithm \ref{alg1}. By omitting the lower order term, the major complexity at each iteration is $\mathcal{O} (N_T^3 )$ which is derived from the calculation of the same matrix inverse, i.e., $(  \delta \mathbf{J}_3 + \tau \mathbf{I}_{N_T} )^{\dagger}$.
Hence, the total complexity of Algorithm \ref{alg1} is $\mathcal{O} ( \upsilon_{max} \log_2(\frac{\tau_u-\tau_l}{\epsilon_1} )  N_T^3 )$, where ${\upsilon}_{max}$ is the maximum iteration number.
\subsection{Convergence Analysis}
In this subsection, we focus on the convergence analysis of the Algorithm \ref{alg1}. Notice that the minorizer function,  $h( \mathbf{w} | \mathbf{w}^{(\upsilon)})$, of $g(\mathbf{w}) $ satisfies the assumptions of MM method, i.e.,
\begin{align}
	g(\mathbf{w}^{(\upsilon)}) = h(\mathbf{x}^{(\upsilon)} | \mathbf{x}^{(\upsilon)}) \leq 
	h(\mathbf{x}^{(\upsilon+1)} | \mathbf{x}^{(\upsilon)}) \leq g(\mathbf{w}^{(\upsilon+1)}), \label{3-27}
\end{align}
where the first equality follows from the assumption (A1), the first inequality is due to the maximum optimization problem \eqref{3-12}, and the second inequality follows from the assumption (A2). \eqref{3-27} implies that the objective function \eqref{2a} is a non-decreasing function of the limit points ${{\bf w}^{(\upsilon)}}$. Moreover, \eqref{2a} has a finite upper bound because of the bounded feasible set of constraint \eqref{2b}. Hence, the sequence ${g({\bf w}^{(\upsilon)})}$ is guaranteed to converge to a finite value. By the following theorem, we will analyze the convergence property of the  suboptimal solution obtained by Algorithm \ref{alg1}.
\begin{theorem} \label{the-1}
	The suboptimal point obtained by Alogorithm \ref{alg1} converges to a KKT point of Problem \eqref{2} for the case of a single communication user who is an extended interference. 
\end{theorem}
\textbf{\textit{Proof: }} Please refer to Appendix \ref{subsec:The-proof-of-the-1}  .  \hspace{10cm} $\blacksquare$
\vspace{-8mm}
\section{Multiple Communication Users} \label{221230c}
In this section, we investigate the scenario of multiple communication users, i.e., $K>1$. 
As discussed in Section \ref{230104a}, 
%
the radar interferences caused by all of the communication users can be modeled as an extended scatterer, i.e., \eqref{extended_C} holds.
Hence, the optimization problem can be formulated as follows
\begin{subequations} \label{4-1}
	\begin{align}
		\max_{\mathbf{W}} \quad  &\log \left[  \det(  L \widetilde{\mathbf{W}} (\mathbf{R}_R + \mathbf{R}_C ) \widetilde{\mathbf{W}}^H   + \sigma_{Z}^2 \mathbf{I}_{KN_R}) \right] - \log \left[  \det( L \widetilde{\mathbf{W}} \mathbf{R}_C \widetilde{\mathbf{W}}^H   + \sigma_{Z}^2 \mathbf{I}_{KN_R})   \right] \label{4-1a} \\
	{\rm s.t.} \quad & {\Vert \mathbf{W} \Vert}_F^2 \leq P_0 \label{4-1b} \\
		&  	R_k \geq r_k, \quad k=1,....,K , \label{4-1c}
	\end{align}
\end{subequations}
where \eqref{4-1c} denotes the transmission rate constraint for each of the communication users.

Note that the objective function \eqref{4-1a} is the same as that in Problem \eqref{2} except the dimension of the beamforming matrix, i.e., $\mathbf{W}$, is $N_T \times K$, where $ K > 1$. Similar to the construction of the surrogate function given in \eqref{3-5}  in  Section \uppercase\expandafter{\romannumeral5},  the surrogate function of \eqref{4-1a} is directly given by
\begin{align}
	2\delta \text{Re}(\tilde{\mathbf{w}}^H \mathbf{j}_1) - \delta^2 \tilde{\mathbf{w}}^H \mathbf{J}_3 \tilde{\mathbf{w}} + \mathrm{const}, \label{4-2}
\end{align}
where $\tilde{\mathbf{w}}$ is defined in Lemma \ref{lemma-2}, and the definitions of the other parameters are the same as \eqref{3-5}.

The power constraint \eqref{4-1b} can be rewritten as
\begin{align}
	\tilde{\mathbf{w}}^H \tilde{\mathbf{w}} \leq P_0.  \label{4-3}
\end{align}

For the purpose of unifying the optimizing variable, the beamforming vector of the $k$-th user can be represented as follows
\begin{align}
	\mathbf{w}_k = ( \mathbf{i}_k^T \otimes \mathbf{I}_{N_T} ) \text{vec}(\mathbf{W}) = ( \mathbf{i}_k^T \otimes \mathbf{I}_{N_T} ) \tilde{\mathbf{w}}, \label{4-4}
\end{align}
where $  \mathbf{i}_k $ is the $k$-th column of the identity matrix $\mathbf{I}_K$.
By the use of the SCA and \eqref{4-4}, the transmission rate constraint in \eqref{rate} can be approximated as follows
\begin{align}
	-2\text{Re}(\mathbf{w}^{(\upsilon),H} \mathbf{L}_k \tilde{\mathbf{w}}) + \mathbf{w}^{(\upsilon),H} \mathbf{L}_k \mathbf{w}^{(\upsilon)} + \nu_k \left( \sum_{j=1,j \neq k}^{K} \tilde{\mathbf{w}}^H \mathbf{L}_{kj} \tilde{\mathbf{w}}   + \sigma_{N}^2  \right) \leq 0,  \quad k=1,....,K , \label{4-5}
\end{align} 
where $\mathbf{L}_k  = ( \mathbf{i}_k \otimes \mathbf{I}_{N_T} )       \mathbf{h}_k \mathbf{h}_k^H    ( \mathbf{i}_k^T \otimes \mathbf{I}_{N_T} ),  \mathbf{L}_{kj} = ( \mathbf{i}_j \otimes \mathbf{I}_{N_T} ) \mathbf{h}_k \mathbf{h}_k^H    ( \mathbf{i}_j^T \otimes \mathbf{I}_{N_T} ) $, and $ \nu_k = 2^{r_k} -1$.

Based on the discussion above, Problem \eqref{4-1} can be approximated as  
\begin{subequations} \label{4-6}
	\begin{align}
		\min_{\tilde{\mathbf{w}}} \quad  &-2 \text{Re}(\tilde{\mathbf{w}}^H \mathbf{j}_1) + \delta \tilde{\mathbf{w}}^H \mathbf{J}_3 \tilde{\mathbf{w}} \label{4-6a} \\
		\mathrm{s.t.} \quad & \eqref{4-3}, \eqref{4-5}, \label{4-6b}
	\end{align}
\end{subequations}
where transformations of  \eqref{4-6a} are the same as that of \eqref{3-12a}. The Problem \eqref{4-6a} is an SOCP problem and the suboptimal solution can be solved using the CVX.
Finally, the proposed Algorithm solving Problem \eqref{4-1} is summarized in Algorithm \ref{alg2}.

\begin{algorithm}[!t]
	\caption{MM-based Algorithm for Solving Problem \eqref{4-1} }
	\label{alg2}
	\begin{algorithmic}[1]
		\STATE Initialize the tolerance $\epsilon_2$, the zero-forcing beamforming matrix $\mathbf{W}^{(0)}$ and the maximum iteration number $\upsilon_{max}$. Set iteration index $ \upsilon = 1 $. Calculate the objective function of Problem \eqref{4-1} $g(\mathbf{W}^{(0)})$.
		\REPEAT
		\STATE Obtain $\mathbf{W}^{(\upsilon)} $ by solving the Problem \eqref{4-6}.
		\STATE Set $ \upsilon = \upsilon + 1 $.
		\UNTIL {$| \frac{g ( \mathbf{W}^{(\upsilon+1)} ) - g ( \mathbf{W}^{(\upsilon)} )}{g ( \mathbf{W}^{(\upsilon)} ) } | \leq \epsilon_2  $}.
	\end{algorithmic}
\end{algorithm} 
\vspace{-5mm}
\subsection{Complexity Analysis}
The complexity of Algorithm \ref{alg2} per iteration is from solving the SOCP Problem \eqref{4-6}. Based on the analysis in \cite{complexity_SOCP} and \cite{multicast}, the complexity of solving an SOCP problem is $\mathcal{O}(NM^{3.5} + N^3 M^{2.5} ) $, where $M$ is the number of SOC constraints and $N$ is the dimension of every SOC constraint. In Problem \eqref{4-6}, there is one power constraint whose dimension is $N_T K$ and $K$ transmission rate constraints, the dimension of each is $N_T K$. Hence, by omitting the lower order terms, the complexity of Algorithm \ref{alg2} per iteration is $\mathcal{O} ( N_TK + N_T^3 K^3 + N_T K^{4.5} + N_T^3 K^{5.5} )$.
\vspace{-5mm}

\subsection{Convergence Analysis}
Similar to the convergence analysis in Section \uppercase\expandafter{\romannumeral5}, the sequence $g(\mathbf{W}^{(\upsilon)})$ can be proved to converge to a limited value.
In addition, the suboptimal solution generated by Algorithm \ref{alg2} converges to a KKT point of Problem \eqref{4-1}.
The proof is similar to Theorem \ref{the-1} and thus, omitted.
\vspace{-5mm}

\section{Numerical Results}
In this section, numerical results are provided to evaluate the performance of the proposed algorithms. 
The DFRC BS is set to have $N_T = N_R = 6$ antennas with the maximum transmit power $P_0 = 40  \,\text{dBm}$.
The number of time slots is $L = 30$, and the noise powers are set as $ \sigma_{N}^2 = 20 \,  \text{dBm} $ and $\sigma_{Z}^2 = 30 \,\text{dBm}$ \cite{CRB}, \cite{beampattern}.
The communication channel $\mathbf{H}$ is an i.i.d. Rayleigh fading channel, where each entry has a complex Gaussian distribution with zero mean and unit variance.
The rate threshold is $r = 6 \,\text{bps/Hz} $ for all users. When considering 
the scenario of multiple communication users, the number of users is set to $K = 3$.
The transmit and receive array steering vectors of the ULA at the DFRC BS are respectively given by
\begin{align}
	\mathbf{a}(\theta) = [ 1, e^{-i \frac{2\pi d_1 }{\lambda} \sin \theta }, \dots, e^{-i \frac{2\pi (N_T - 1)d_1 }{\lambda} \sin \theta }  ], \thinspace  \text{and} \thinspace
	\mathbf{b}(\theta) = [ 1, e^{-i \frac{2\pi d_2 }{\lambda} \sin \theta }, \dots, e^{-i \frac{2\pi (N_R - 1)d_2 }{\lambda} \sin \theta }  ], \nonumber
\end{align}
where $\lambda$ is the wavelength, and $d_1$ and $d_2$ denote the transmit and receive antenna spacing, respectively. Here, we consider $ d_1 = d_2 = \frac{\lambda}{2}$.
The desired point target to be sensed is assumed to be  located at $\theta^r = 0^{\circ}$ and the
average strength of the echo signal is given as  $ \beta^2 =1$. 
In the case of a single communication user, it is placed
at $\theta^c = -30^{\circ}$ and the average strength of the interfering echo signal generated by the communication user is set as  $ \gamma^2 = 100$. 
In the multi-user scenario, users are assumed to be located in the angle range of  $[-30^{\circ}, -25^{\circ}]$, forming an extended scatterer. We assume that the extended scatterer comprises $M=50$ point-like scatterers with angle $\theta_{i}^{c} \in [-30^{\circ}, -25^{\circ}]$ and average strength $\gamma_i^2 = 100$.
Moreover, maximum ratio transmission beamforming and zero-forcing beamforming are set as the initial points of ISAC beamforming in 
Algorithms 1 and 2, respectively.
%
%
Finally, the stopping criteria of Algorithms \ref{alg1} and \ref{alg2} are set to $ \epsilon_1 = 10^{-8}$ and $\epsilon_2 = 10^{-6}$, respectively.

The schemes considered in this work and benchmark schemes are listed as follows: 
``W/O-inter-SU'' represents the scheme given in Section \ref{221230} for the case a single communication user that does not produce any echo interference; 
``W-point-inter-SU'' denotes the scheme given in Section \ref{point-inter} with a point echo interference from a single communication user; 
``W-extended-inter-SU'' denotes the scheme of Algorithm \ref{alg1}, which is for the case of a single communication user who is an extended interference; and 
``W-extended-inter-MU'' represents the scheme of Algorithm \ref{alg2}, which is the scenario where the radar sensing suffers from an extended echo interference  generated by multiple communication users. For performance comparison with existing results, we also consider the performance of Algorithm \ref{alg2}, under the setting of $\mathbf{R}_C = 0$, i.e., though there are multiple communication users, they do not generate any echo interference. This scheme is denoted by ``W/O-inter-MU''. 

Our results are compared to that of existing schemes such as
the ISAC beamforming based on the CRB metric \cite{CRB} and beampattern metric \cite{beampattern}.   “CRB-SU” and “Beampattern-SU” denote the  schemes with single communication user, and “CRB-MU” and “Beampattern-MU” denote the schemes with multiple communication users. 
Since the algorithms proposed in \cite{CRB} and \cite{beampattern} do not consider the echo interference  from communication users, here we only compare the performance of MI, CRB, and beampattern metrics in the scene without echo interference.
\vspace{-5mm}

\subsection{Convergence and complexity of the algorithms}
Table \uppercase\expandafter{\romannumeral1} compares the computational time consumed when running the proposed different algorithms until convergence. \vspace{-5mm}
\begin{table}[htb]     
	\begin{center}   
		\caption{The comparision of computational time.}  
		\label{table1} 
		\centering
		\begin{tabular}{|m{9cm}|m{3cm}|}   
			\hline   \textbf{Algorithms} & \textbf{Run-Time (s)}    \\
			\hline   W/O-inter-SU based on the closed-form solution to Problem \eqref{1} & 0.009  \\ 
			\hline   W/O-inter-SU based on the SDR method for Problem \eqref{1} & 0.62   \\  
			\hline   W-point-inter-SU based on the SDR method for Problem \eqref{2} \newline with a single communication user who is a point interference&  0.702  \\  
			\hline   W-extended-inter-SU based on Algorithm \ref{alg1} for Problem \eqref{2} \newline with a single communication user who is an extended interference & 13.451   \\     
			\hline   W-extended-inter-SU based on SOCP for Problem \eqref{2} \newline with a single communication user who is an extended interference  &  942.2 \\ 
			\hline   W-extended-inter-MU based on Algorithm \ref{alg2} for Problem \eqref{4-1} &  407.942  \\ 
			\hline   
		\end{tabular}   
	\end{center}   
\end{table} \vspace{-10mm}

For Problem  \eqref{1}, while the performance of the closed-form solution and that of the SDR method is the same, it can be seen from Table \ref{table1} that, the computational time of the closed-form solution is much faster than that of the SDR method. This shows that the proposed closed-form solution to Problem \eqref{1} has a lower complexity compared to the conventional SDR method based on the Gaussian randomization. 

\begin{figure}[htbp]
	\centering \includegraphics[width=3.4in,height=2.6in]{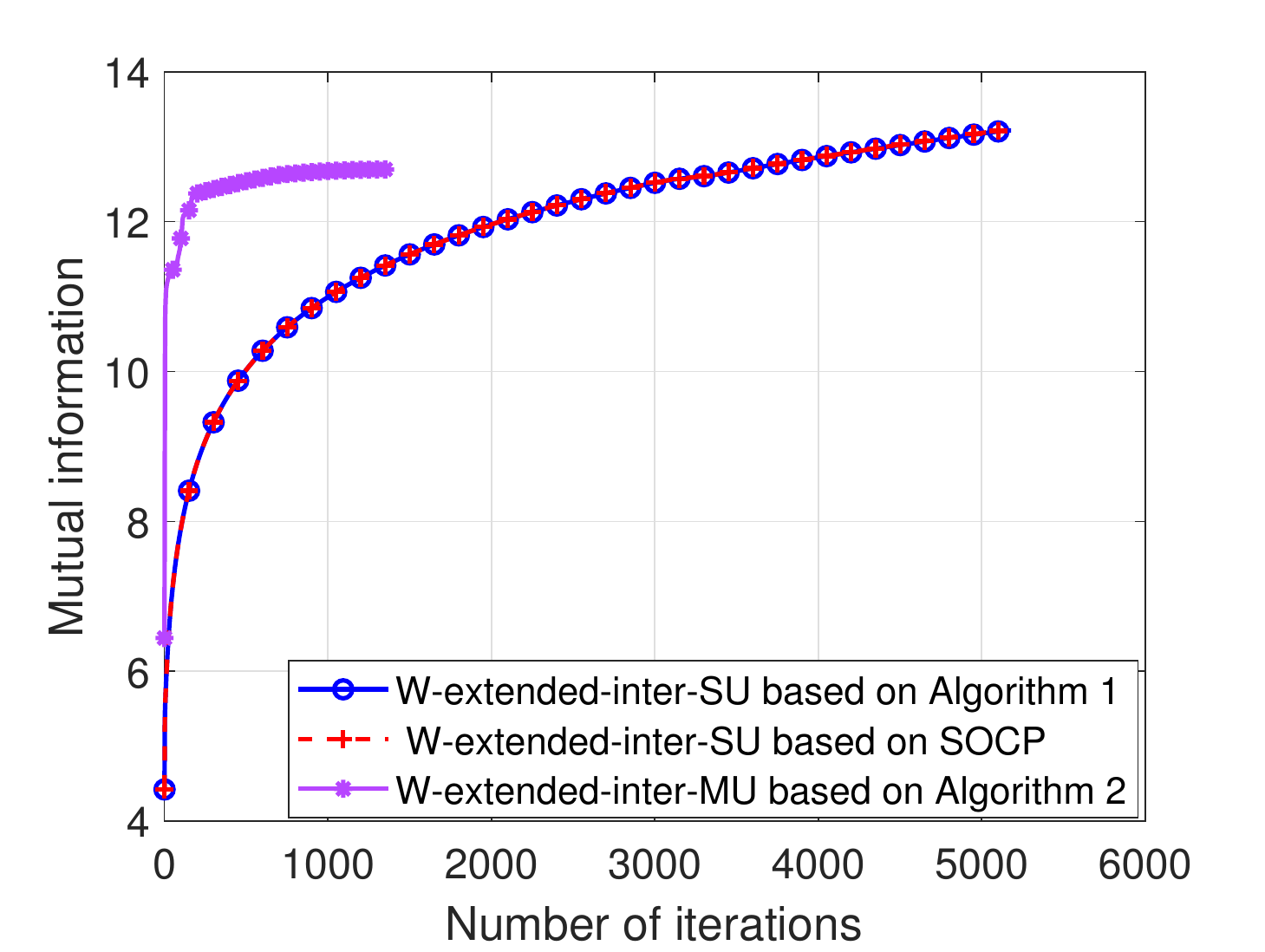}
	\caption{Mutual information versus number of iterations.}
	\label{iteration} 
\end{figure}

For Problem (\ref{2}) with a single communication user who is an extended interference, we plot the achieved value of the cost function vs. the number of iterations in Fig. \ref{iteration}. 
%
%
``W-extended-inter-SU based on SOCP'' denotes the same problem solved by SOCP. As can be seen, to achieve the same value of the cost function, the number of iterations needed by Algorithm \ref{alg1}  is almost the same number as that of the SOCP-based algorithm. However, as is shown in the Table \ref{table1}, Algorithm \ref{alg1} only needs $13.452 \,\text{s}$ to converge while the SOCP-based algorithm needs $942.2 \,\text{s}$ to converge. This illustrates that the computational complexity per iteration of Algorithm \ref{alg1} is much less than that of the SOCP per iteration. 

For Problem \eqref{4-1}, though Algorithm \ref{alg2} converges within $1200$ iterations as shown in Fig. \ref{iteration}, according to Table \ref{table1}, it requires about $t=408\,\text{s}$  of computational time. This is due to the high complexity of the SOCP that needs to be solved in each iteration. 
\vspace{-5mm}

\subsection{The single communication user scenario}
In this subsection, 
we discuss the scenarios with or without echo interference from the single communication user, which are delineated in Sections \uppercase\expandafter{\romannumeral3}, \uppercase\expandafter{\romannumeral4} and \uppercase\expandafter{\romannumeral5}. 

\begin{figure}[htbp]
	\centering
	\begin{minipage}[t]{0.48\textwidth}
		\centering
		\hspace*{-5mm}\includegraphics[width=3.4in,height=2.6in]{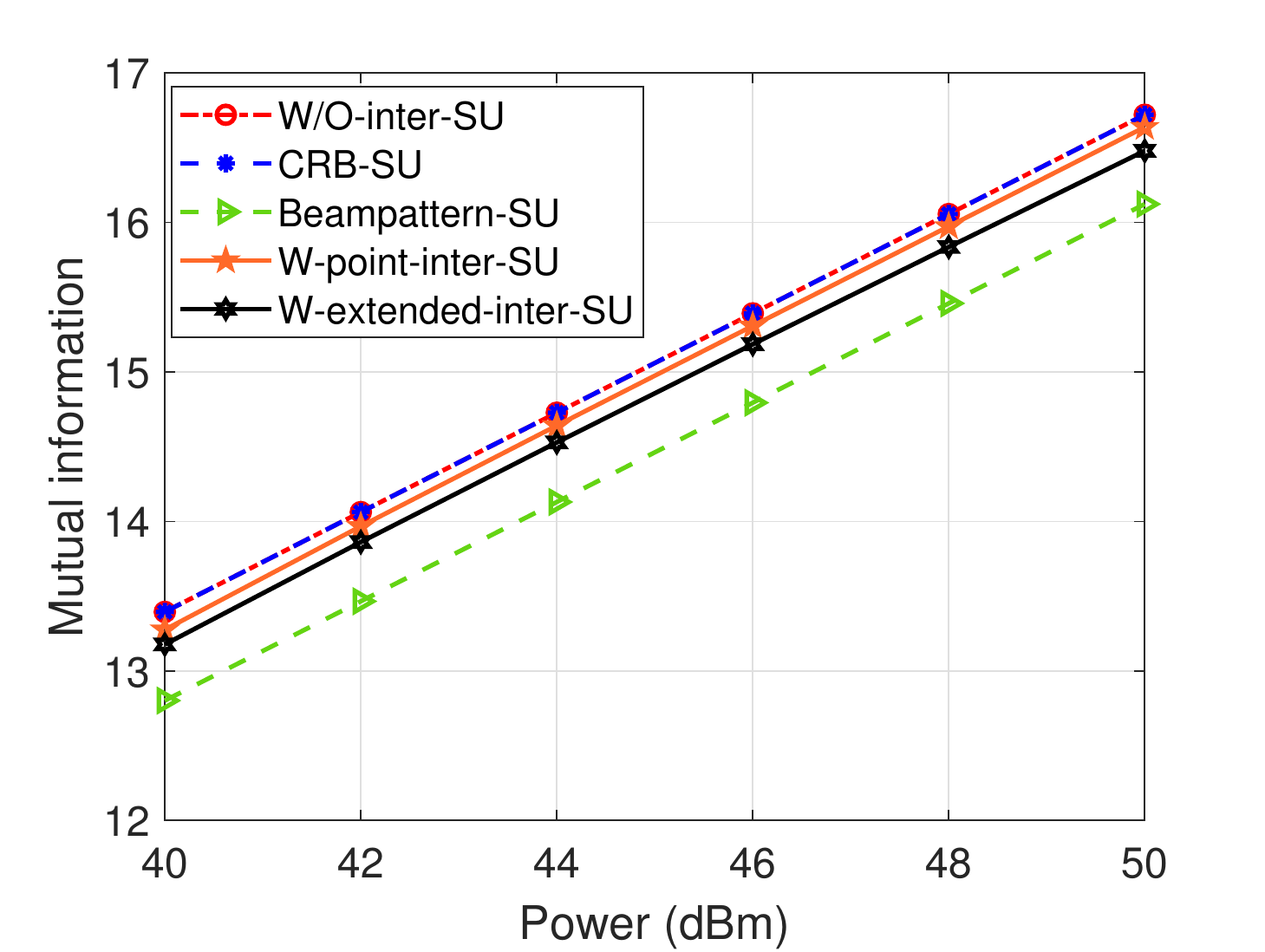}
		\caption{Achievable MI versus maximum transmit power.}
		\label{power-mi} 
	\end{minipage}\hspace*{8mm}
	\begin{minipage}[t]{0.48\textwidth}
		\centering 
		\hspace*{-5mm}\includegraphics[width=3.4in,height=2.6in]{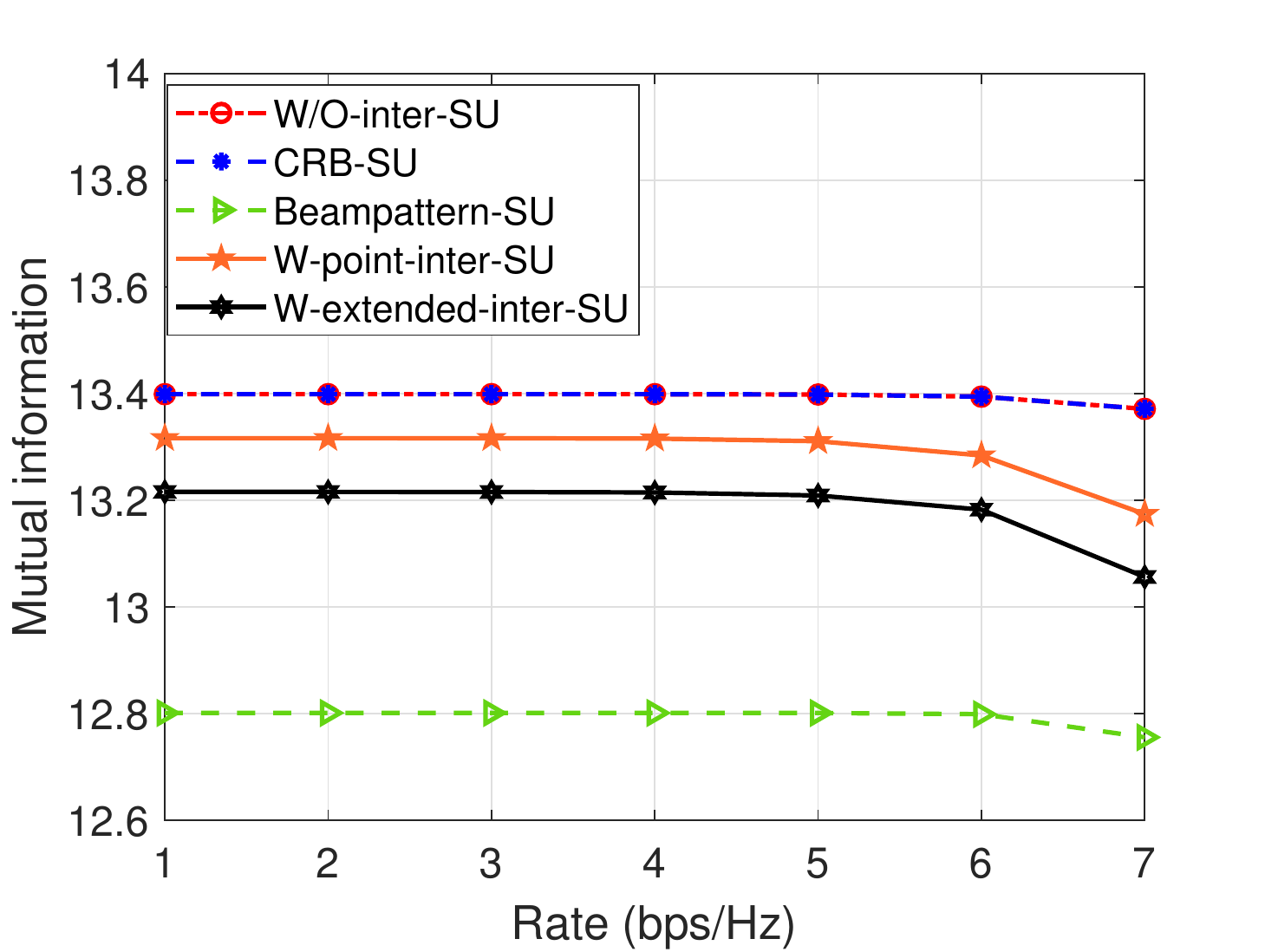}
		\caption{Achievable MI versus target communication rate.}
		\label{rate-mi} 
	\end{minipage}
\end{figure}

Fig. \ref{power-mi} plots the MI as a function of the maximum transmit power, $P_0$. We make the following observations. 
First, it is obvious that by consuming more transmit power, 
a higher MI can be attained, which leads to  more accurate radar sensing performance. 
Second,  in the scheme without echo interference,
the achievable MI of Problem \eqref{1} based on the MI metric, studied in this paper,  is the same as that of the scheme based on CRB metric \cite{CRB} and higher than that  of the scheme based on beampattern metric \cite{beampattern}.
This is because the optimization problem with the MI metric in this case is the same as that with  the CRB metric.
Furthermore, beampattern metric is the transmit characteristic of the BS, which may result in a lower sensing performance than that achieved based on receive performance characteristic, i.e., MI and CRB metrics.
%
%
Third, it is observed that different types of echo interferences from the single communication user  have different effects on radar sensing performance. More specifically, 
compared with the scenario with no echo interference, the echo interference from the point scatterer only produces a slight loss in MI performance, while the echo interference from the extended scatterer causes a large performance loss. 
This is because the total average strength of the interfering echo signals from the extended scatterer is large.

%
%
%
Fig. \ref{rate-mi} depicts the achievable MI versus the target communication rate. 
It is observed that when $r \leq
6$ bps/Hz, the ISAC beamforming can both meet the QoS requirement of the communication user and achieve the best radar sensing performance. However, when $r$ is increased to $7$ bps/Hz,  the radar sensing performance deteriorates in order to  satisfy  the  QoS requirement of the communication user. 
%
%
%
In this case, there exists a tradeoff between radar sensing and communication.
In addition, within the range of radar-communication tradeoff, i.e., $r >
6$ bps/Hz, the echo interference  from the single communication user causes a sharper performance loss on MI, compared to case with no echo interference. 
Therefore, the echo interference from the communication user cannot be ignored when designing the ISAC beamforming, because it may  cause significant performance loss to radar sensing.


\begin{figure}[htbp]
	\begin{minipage}[t]{0.48\textwidth}
		\centering     \hspace*{-5mm}\includegraphics[width=3.4in,height=2.6in]{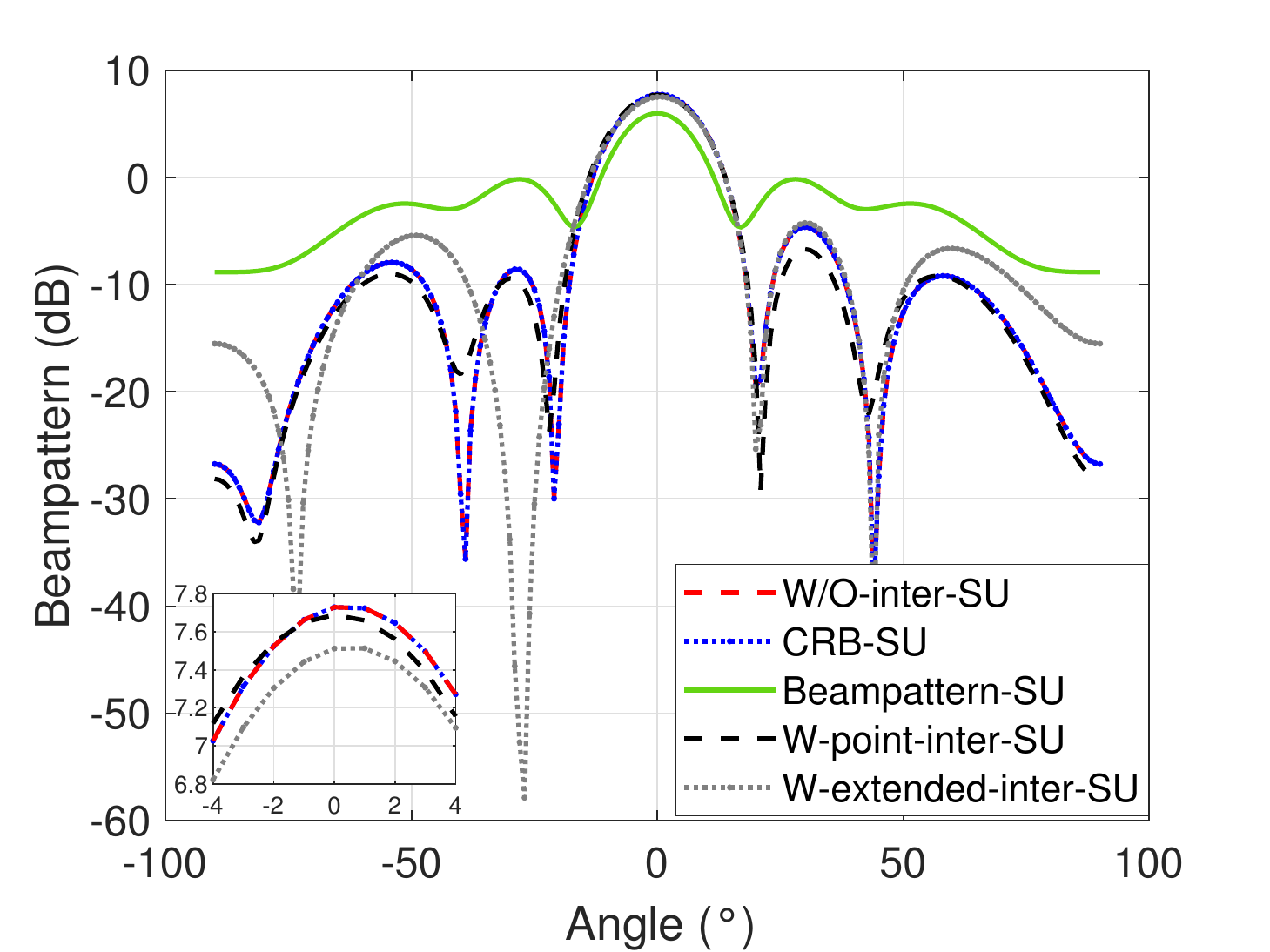}
		\caption{Beampatterns in the scenarios of single communication user.}
		\label{beampattern-su} 
	\end{minipage}\hspace*{8mm}
	\begin{minipage}[t]{0.48\textwidth}
		\centering     \hspace*{-5mm} \includegraphics[width=3.4in,height=2.6in]{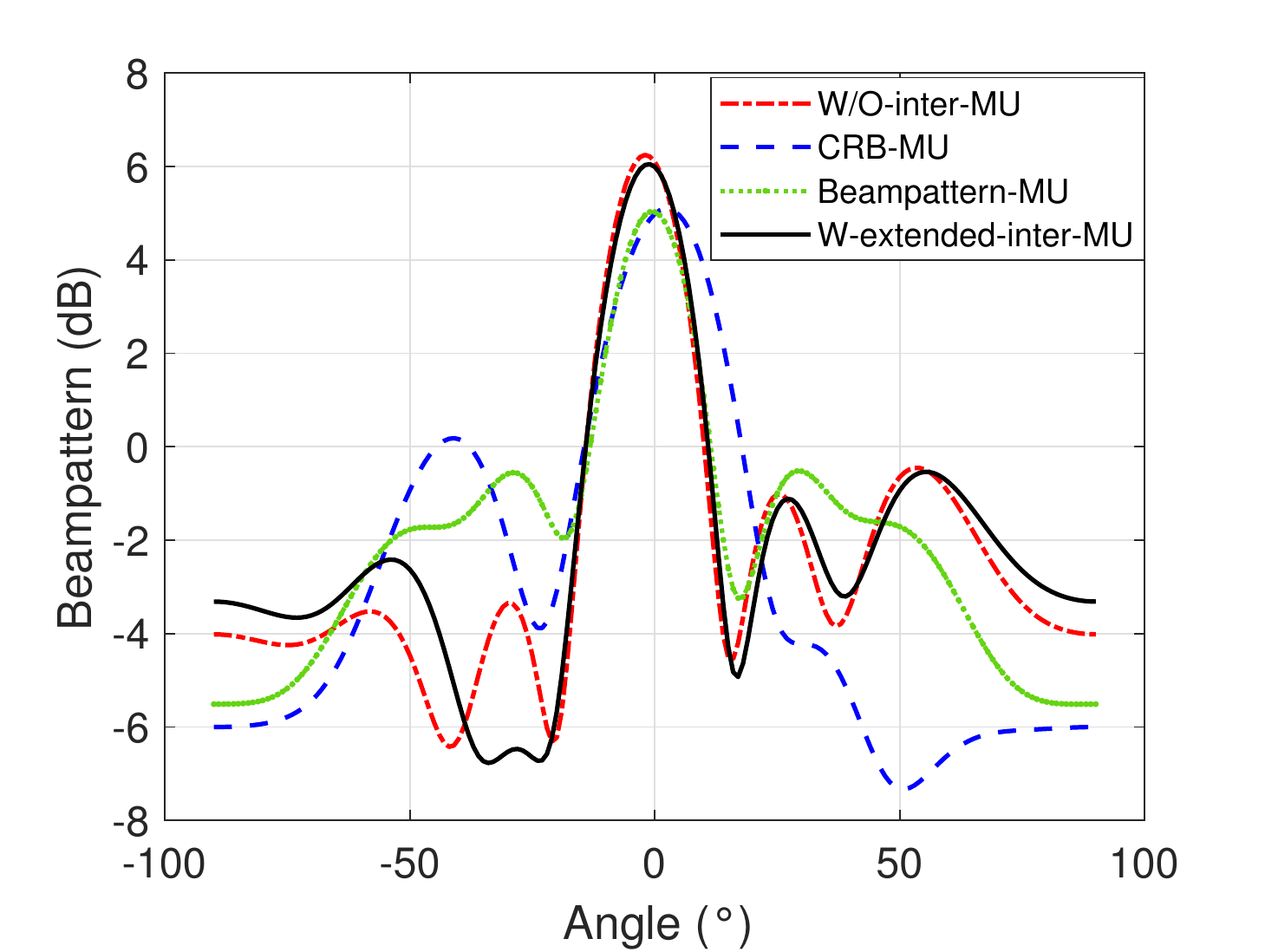}
		\caption{Beampatterns in the scenarios of multiple communication users.}
		\label{beampattern-mu} 
	\end{minipage}
\end{figure}

As explained before, beampattern is a common transmit characteristic illustrating the performance of  radar sensing \cite{lijian}.
Fig. \ref{beampattern-su} shows the beampatterns of the radar target for different schemes in the scenario of a single communication user. 
Firstly, we focus on the radar beampatterns of the schemes without echo interference.
Consistent with the conclusion drawn from Fig. \ref{power-mi}, the ISAC beamforming designed based on the MI metric and the CRB metric can achieve the same beampattern. Compared to the beambattern designed based on the beampattern metric \cite{beampattern}, 
the mainlobe of the beampattern based on the MI metric is higher, and furthermore, it is more effective in suppressing  the sidelobe. 
%
%
This reveals performance superiority in terms of radar sensing based on the MI metric over the beampattern metric.  
%
%
%
Secondly, we observe that
the beampatterns have almost the same mainlobe at the radar target location of $0^{\circ}$, with a gap of about 0.2 dB at most, for the schemes without echo interference, with the point echo interference and with the extended echo interference. Moreover, at the location of the communication user, i.e., $[-30^{\circ}, -25^{\circ} ]$, the suppression is remarkable for the case of the extended echo interference. This shows that optimizing MI can suppress the echo interference effectively.  
\vspace{-5mm}

\subsection{The multiple communication users scenario}

In this subsection, we study the case of multiple communication users with or without the
extended echo interference.

Fig. \ref{beampattern-mu} compares the beampatterns of different schemes.
With the increase of the number of communication users, the differences of the beampatterns in the case of no echo interference for the CRB \cite{CRB}, MI and beampattern \cite{beampattern} metrics  have become larger, compared to Fig. \ref{beampattern-su}. The ISAC beamforming scheme based on the MI metric, i.e., Algorithm\ref{alg2}, under the setting of $\mathbf{R}_C = 0$,  achieves the best beampattern. This illustrates that optimizing the MI metric is beneficial to achieving a better beampattern for radar sensing. In addition, the extended echo interference from the directions of  $[-30^{\circ}, -25^{\circ} ]$ is suppressed considerably, which verifies again that optimizing MI can suppress the echo interference effectively. 


\begin{figure}[htbp]
	\centering \includegraphics[width=3.4in,height=2.6in]{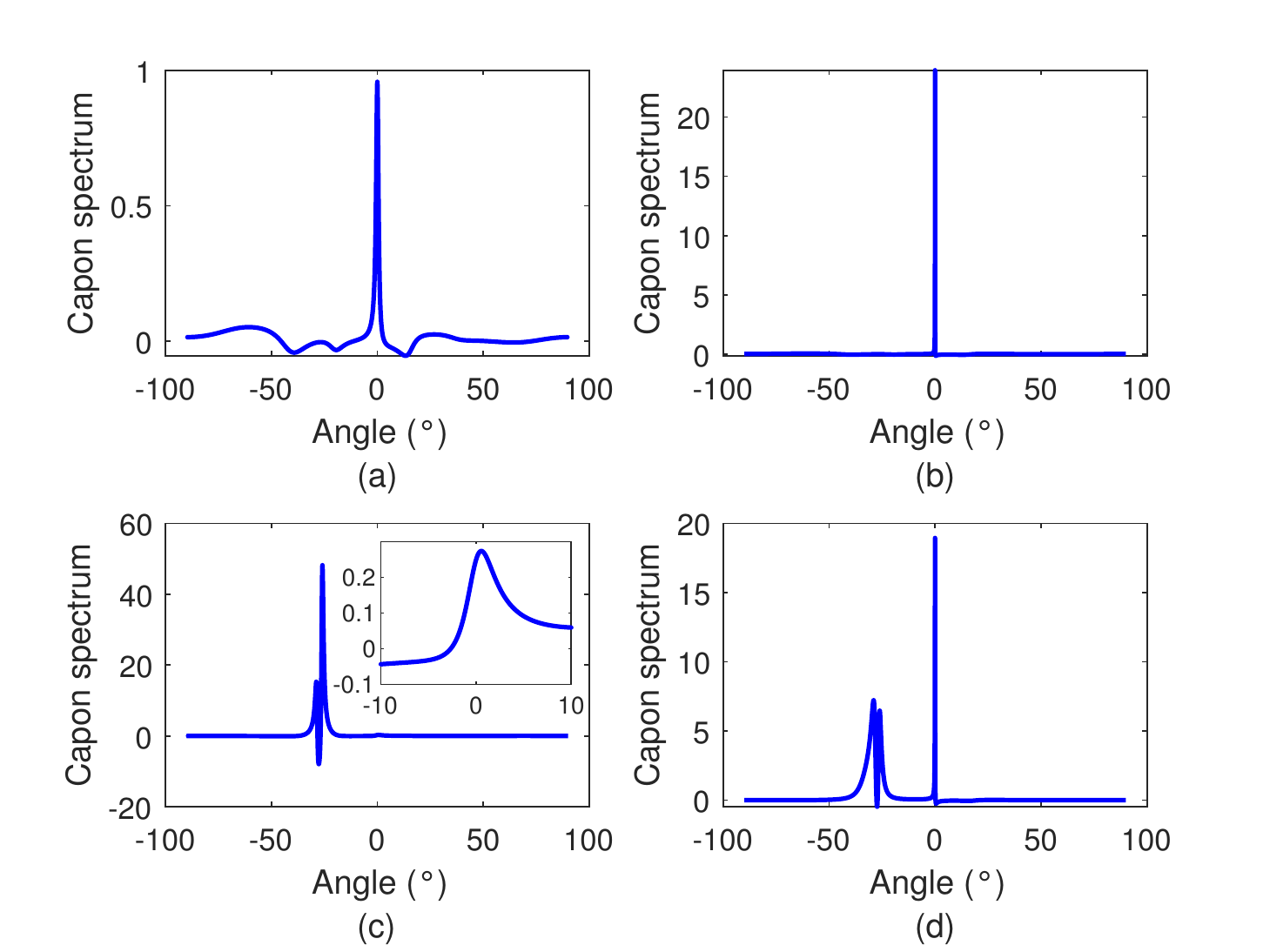}
	\caption{Capon spatial spectrum. (a) Capon spatial spectrum without echo interference and the average strength is $\beta^2 = 1$. (b) Capon spatial spectrum without echo interference and the average strength is $\beta^2 = 25$. (c) Capon spatial spectrum with the extended echo interference from the communication and the average strength are  $\beta^2 = 1, \gamma^2 = 100$. (d) Capon spatial spectrum with the extended echo interference from the communication and the average strength are $\beta^2 = 25, \gamma^2 = 1$. }
	\label{capon-spectrum} 
\end{figure}
For the purpose of comparing angle estimation performance between W/O-inter-MU and W-extended-inter-MU, 
Fig. \ref{capon-spectrum} exhibits the capon spatial spectrum with and without the extended echo interference in different average strengths \cite{capon}. It is observed from Fig. \ref{capon-spectrum} (a) and (c) that the  angle estimation accuracy of the desired target is severely affected by strong echo interference at the location of $[-30^{\circ}, -25^{\circ}]$. However, from the comparison between Fig. \ref{capon-spectrum} (b) and (d), the peak at the location of $0^{\circ}$ in Fig. \ref{capon-spectrum} (d) is also prominent when there exists  weak echo interference. 
This shows that ISAC beamforming designed based on the MI metric can suppress the weak echo interference efficiently, but cannot guarantee accurate angle estimation in the existence of strong echo interference. 



\begin{figure}[htbp]
	\centering \includegraphics[width=3.4in,height=2.6in]{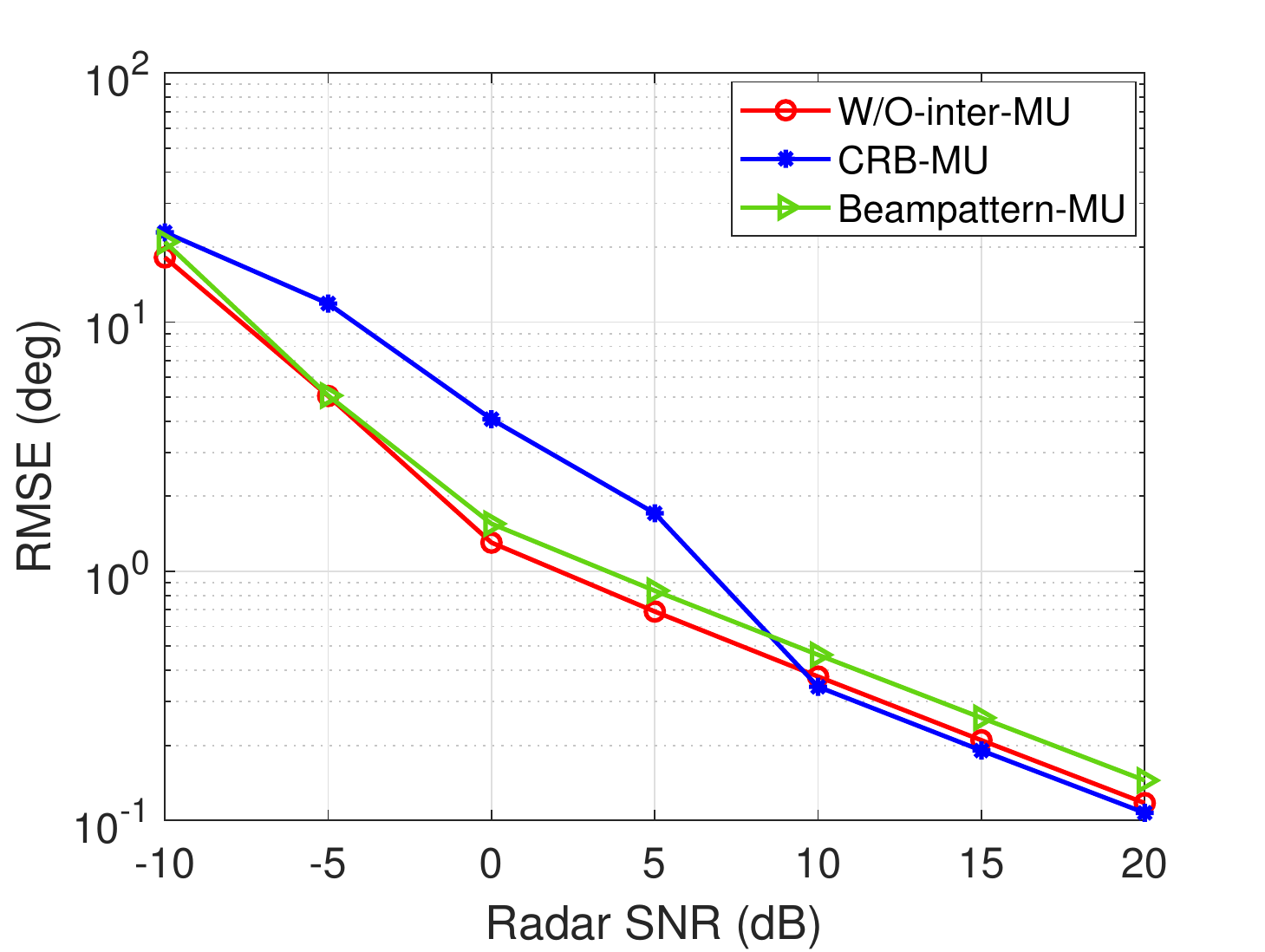}
	\caption{RMSE in the scenario of multiple communication users.}
	\label{RMSE-mu} 
\end{figure}
In order to further illustrate the advantages of the MI metric, Fig. \ref{RMSE-mu} shows the RMSE of target angle estimation for different radar performance metrics versus the radar SNR which is defined as $ 10 \log (\frac{\beta^2 L P_0}{\sigma_{Z}^2}) $ \cite{CRB},\cite{MLE}. Here, we set $r_k = 4 \,\text{bps/Hz}$ for all users,
and use  the maximum likelihood estimation (MLE) to estimate the angle of the desired radar target \cite{MLE}. 
It is observed that when radar SNR is large, 
the RMSE of the scheme based on the MI metric is better than that of the scheme based on the beampattern metric \cite{beampattern}, but slightly worse than that of the scheme based on the CRB metric \cite{CRB}.
By contrast, at low radar SNR (below 10 dB), the RMSE of the scheme based on the MI metric outperforms the other two schemes.
This further shows that the ISAC beamforming scheme based on the MI metric can result in better radar sensing performance.
\vspace{-5mm}

\section{conclusions}
In this paper, we investigated the ISAC beamforming design of the DFRC BS based on the MI metric. Our aim is to maximize the MI of sensing while ensuring the QoS of each communication user. Moreover, the interfering echo signals from the communication users were considered.
In the scenario of a single communication user,  three types of echo interference  were studied: no interference, the point echo interference and the extended echo interference. 
A closed-form solution, an SDR-based method, and a low-complexity algorithm based on the MM method were proposed, respectively.
In the scenario of multiple communication users, 
the ISAC beamforming problem was studied under the extended echo interference. This problem was solved by an MM-based algorithm.
Numerical results demonstrated that the proposed ISAC beamforming scheme designed based on the MI metric achieves better beampattern and RMSE of angle estimation compared to schemes based on the CRB and the beampattern metrics. Moreover, maximizing MI can suppress the echo interference remarkably.

\appendices{}   
\section{ THE PROOF OF LEMMA \ref{lemma-1} \label{proof1}   }

The power constraint \eqref{1b} must be satisfied with equality for the optimal solution $\mathbf{\hat{w}}$. We prove this by contradiction. 
Assume that the optimal solution is $\mathbf{\hat{w}}_1$ with the power $\mathbf{\hat{w}}_1^H \mathbf{\hat{w}}_1 = P_1 < P_0 $. Then, $\mathbf{\hat{w}}_2 = \sqrt{\frac{P_0}{P_1}} \mathbf{\hat{w}}_1$ is a feasible solution because it satisfies the constraints \eqref{1b} and \eqref{single SINR}, i.e., 
\begin{align}
	&\mathbf{\hat{w}}_2^H \mathbf{\hat{w}}_2 = \frac{P_0}{P_1} \mathbf{\hat{w}}_1^H \mathbf{\hat{w}}_1 = P_0, \nonumber \\
	&{\lvert \mathbf{h}^H \mathbf{\hat{w}}_2 \rvert}^2 = \frac{P_0}{P_1}{\lvert \mathbf{h}^H \mathbf{\hat{w}}_1 \rvert}^2 > {\lvert \mathbf{h}^H \mathbf{\hat{w}}_1 \rvert}^2 \geq \Omega.
\end{align}
Note that the feasible solution $\mathbf{\hat{w}}_2$ has a larger objective value than $\mathbf{\hat{w}}_1$, while $\mathbf{\hat{w}}_2$ satisfies the constraints \eqref{1b} and \eqref{single SINR}. This contradicts the optimality of $\mathbf{\hat{w}}_1$. Therefore, the optimal solution $\mathbf{\hat{w}}$ must satisfy the power constraint \eqref{1b} with equality. 

Hence, we can reformulate the problem \eqref{1-2} as follows
\begin{subequations}\label{app1}
	\begin{align} 
		\max_{\mathbf{w}} \quad  &  \mathbf{w}^H \mathbf{a}(\theta)  \mathbf{a}^H(\theta) \mathbf{w}  \label{app1a} \\
		\text{s.t.} \quad & \mathbf{w}^H \mathbf{w} = P_0  \label{app1b} \\
		&{\lvert \mathbf{h}^H \mathbf{w} \rvert}^2 \geq \Omega. \label{app1c}
	\end{align}
\end{subequations}
We discuss the two cases of the SNR constraint of \eqref{app1c}. If the optimal $\mathbf{w}$ satsifies (\ref{app1c}) strictly, then we only need to consider the power constraint \eqref{app1b}. It can be easily verified that the optimal solution to \eqref{app1} is $\mathbf{\hat{w}} = \sqrt{P_0} \frac{\mathbf{a}(\theta)}{\Vert \mathbf{a}(\theta) \Vert}$. This case happens when ${\lvert \mathbf{h}^H \mathbf{\hat{w}} \rvert}^2 > \Omega$, i.e., ${\lvert \mathbf{h}^H \mathbf{a}(\theta) \rvert}^2 > \frac{ \Omega {\Vert \mathbf{a}(\theta)  \Vert}^2 }{P_0} $. On the other hand, if the optimal $\mathbf{w}$ satsifies (\ref{app1c}) with equality, then 
we may normalize the parameters in the problem \eqref{app1} by taking $\mathbf{c} = \frac{\mathbf{w}}{\Vert \mathbf{w}  \Vert} , \mathbf{v}_2 = \frac{\mathbf{a}(\theta)}{\Vert \mathbf{a}(\theta)  \Vert}  , \mathbf{v}_1 = \frac{\mathbf{h}}{\Vert \mathbf{h}  \Vert}$, and $t = \frac{\Omega}{P_0 {\Vert \mathbf{h}  \Vert}^2}$. The problem can be reformulated as
\begin{subequations} \label{app3}
	\begin{align}
		\max_{\mathbf{c}} \quad  &  \mathbf{c}^H \mathbf{v}_2  \mathbf{v}_2^H \mathbf{c}  \label{app3a}\\
		\text{s.t.} \quad & \mathbf{c}^H \mathbf{c} = 1  \label{app3b} \\
		& \mathbf{c}^H \mathbf{v}_1  \mathbf{v}_1^H \mathbf{c}  = t  \label{app3c},
	\end{align}
\end{subequations}
Note that from Cauchy-Schwarz inequality, we have $ 0 \leq t \leq 1 $.
By applying Lemma 2 in \cite{proof1} and $\mathbf{c} = \frac{\mathbf{w}}{\Vert \mathbf{w}  \Vert}$, the optimal solution $ \mathbf{\hat{w}}$ is given by
\begin{align*}
	\begin{split}
		\mathbf{\hat{w}} = \left \{
		\begin{array}{lr}
			\sqrt{P_0} \frac{\mathbf{h} }{\Vert \mathbf{h}  \Vert},     &  \frac{\lvert \mathbf{a}^H(\theta) \mathbf{h} \rvert}{\Vert \mathbf{a}(\theta)  \Vert \Vert \mathbf{h}  \Vert } =  1 \\
			z_1 \frac{\mathbf{h}}{\Vert \mathbf{h} \Vert} + z_2 \frac{\mathbf{a}(\theta)}{\Vert \mathbf{a}(\theta) \Vert},                               &  0 \leq \frac{\lvert \mathbf{a}^H(\theta) \mathbf{h} \rvert}{\Vert \mathbf{a}(\theta)  \Vert \Vert \mathbf{h}  \Vert } <  1
		\end{array}
		\right.
	\end{split}
\end{align*}
where $z_1 = \sqrt{P_0} ( \sqrt{t} - u_2 r ) \frac{\mathbf{a}^T \mathbf{h}^*}{| \mathbf{a}^H \mathbf{h} | }, z_2 = \sqrt{\frac{P_0 (1-t) }{1-r^2}}, t =  \frac{\Omega}{P_0 {\Vert \mathbf{h}  \Vert}^2}, r = \frac{{\lvert \mathbf{a}^H(\theta) \mathbf{h} \rvert} }{  {\Vert \mathbf{a}(\theta)  \Vert} {\Vert \mathbf{h}  \Vert} }$,  and $u_2 = \sqrt{\frac{1-t}{1-r^2}}$.

Hence, the proof is completed.
\vspace{-5mm}

\section{ THE PROOF OF LEMMA \ref{lemma-2} \label{proof2}  } 
In this lemma, we
investigate the relationship between $\tilde{\mathbf{w}}$ and $\text{vec}(\widetilde{\mathbf{W}})$ for $K \geq 1$.
Particularly, we have 
\begin{align}
	\widetilde{\mathbf{W}} &= \mathbf{I}_{N_R} \otimes \mathbf{W}^H = \left( \mathbf{I}_{N_R} \otimes \mathbf{W}^H  \right) \mathbf{I}_{N_T N_R} 
	= \left( \mathbf{I}_{N_R} \otimes \mathbf{W}^H  \right) \left[  \mathbf{c}_1, \mathbf{c}_2,...,\mathbf{c}_{N_T N_R}  \right] ,
\end{align}
where $\mathbf{c}_i \in \mathbb{R}^{N_TN_R \times 1} , i =1,2,...,N_T N_R $ is a unit vector whose $i$-th element is one and the other elements are all zero. Following from the property of the Kronecker product, we can obtain that 
\begin{align}
	\left( \mathbf{I}_{N_R} \otimes \mathbf{W}^H  \right) \mathbf{c}_i &= \text{vec} \left(  \mathbf{W}^H  \mathbf{C}_i \right) 
	= \left(  \mathbf{C}_i^T  \otimes \mathbf{I}_K \right) \mathbf{K}_{N_TK}        \tilde{\mathbf{w}}^*  ,
\end{align}
where $\mathbf{K}_{N_TK} $ is the commutation matrix and $\text{vec}\left(  \mathbf{C}_i  \right)  =   \mathbf{c}_i $. Thus, $\text{vec}( \widetilde{\mathbf{W}} )$ can be represented as follows
\begin{align}
	\text{vec}( \widetilde{\mathbf{W}} ) = \left[   
	\mathbf{C}_{1} \otimes \mathbf{I}_K,   
	\mathbf{C}_{2} \otimes \mathbf{I}_K,   
	\dots ,
	\mathbf{C}_{N_TN_R} \otimes \mathbf{I}_K  
	\right]^T \mathbf{K}_{N_TK} \tilde{\mathbf{w}}^*= \mathbf{F} \tilde{\mathbf{w}}^*,
\end{align}
where $\mathbf{F} =\left[   
\mathbf{C}_{1} \otimes \mathbf{I}_K,   
\mathbf{C}_{2} \otimes \mathbf{I}_K,   
\dots ,
\mathbf{C}_{N_TN_R} \otimes \mathbf{I}_K  
\right]^T \mathbf{K}_{N_TK}$. 

Hence, the proof is completed.
\vspace{-5mm}

\section{ THE PROOF OF LEMMA \ref{lemma-3} \label{proof3} } 
To ensure the feasibility of Problem \eqref{2} for the case of a single communication user who is an extended interference, we formulate Problem \eqref{3-13}.
Similar to the proof in Appendix \ref{proof1}, the optimal value of Problem \eqref{3-13}, $\Omega_1$, is obtained when the power constraint \eqref{3-13b}  is achieved with equality. Thus, Problem \eqref{3-13} is equivalent to the following problem
\begin{subequations}\label{appc-3}
	\begin{align}
		\max_{\mathbf{w}} \quad  &  \mathbf{w}^H \mathbf{h} \mathbf{h}^H \mathbf{w}  \label{appc-3a} \\
		{\rm s.t.} \quad & \mathbf{w}^H \mathbf{w} = P_0.  \label{appc-3b} 
	\end{align}
\end{subequations}

From the physical meaning of the problem, it is easy to find that the optimal value $\Omega_1$ is an increasing function of the maximum transmit power $P_0$. Evidently, if $\Omega_1 > \Omega$, there must exist a feasible solution, $\mathbf{\hat{w}}_3$, to Problem \eqref{3-13} which satisfies the conditions: $\Omega < \mathbf{\hat{w}}_3^H \mathbf{h} \mathbf{h}^H \mathbf{\hat{w}}_3 < \Omega_1$ and $\mathbf{\hat{w}}_3^H \mathbf{\hat{w}}_3 < P_0$.
Hence, the Slater's condition of Problem \eqref{2} for the case of a single communication user who is an extended interference, is satisfied with the strictly feasible solution $\mathbf{\hat{w}}_3$.

Next, we discuss the relationship between the feasibility of Problem \eqref{2} for the case of a single communication user who is an extended interference,  and Problem \eqref{3-12}. The feasibility problem of Problem \eqref{3-12} is given by
\begin{subequations}\label{appc-4}
	\begin{align}
		\max_{\mathbf{w}} \quad  &   2 \text{Re} \left(   \mathbf{w}^{(\upsilon) H} \mathbf{h} \mathbf{h}^H \mathbf{w}  \right)  - \mathbf{w}^{(\upsilon) H} \mathbf{h} \mathbf{h}^H \mathbf{w}^{(\upsilon)} \label{appc-4a} \\
		{\rm s.t.} \quad & \mathbf{w}^H \mathbf{w} \leq P_0.  \label{appc-4b} 
	\end{align}
\end{subequations}

Compared to Problem \eqref{3-13}, Problem \eqref{appc-4} is relaxed with the SCA method and would converge to the same objective function. So, if Problem \eqref{2} is feasible, Problem \eqref{3-12} is also feasible.
Due to this, $\mathbf{\hat{w}}_3$ is also a strictly feasible solution to Problem \eqref{3-12}, that is, the Slater's condition of Problem \eqref{3-12} is satisfied. 

Hence, the proof is completed.
\vspace{-5mm}

\section{ THE PROOF OF THEOREM \ref{the-1} \label{subsec:The-proof-of-the-1}  } 
Supposing $\mathbf{w}^o$ is the suboptimal point generated by the Algorithm \ref{alg1}.
Note that minimizing $- h(\mathbf{w} | \mathbf{w}^{(\upsilon)})$ is equivalent to minimize the objective function of Problem \eqref{3-12}, i.e., $\mathbf{w}^o$ is also the suboptimal point for the problem that minimizes $- h(\mathbf{w} | \mathbf{w}^{(\upsilon)})$ subject to constraints \eqref{3-12b} and \eqref{3-12c}. 
Let $h_1 (\mathbf{w} | \mathbf{w}^{(\upsilon)}) =
- 2 \text{Re} \left(   \mathbf{w}^{(\upsilon),H} \mathbf{h} \mathbf{h}^H \mathbf{w}  \right)  + \mathbf{w}^{(\upsilon),H} \mathbf{h} \mathbf{h}^H \mathbf{w}^{(\upsilon)} $ and $ g_1 (\mathbf{w}) = - \mathbf{w}^H \mathbf{h} \mathbf{h}^H \mathbf{w}$.
From the optimality condition \eqref{3-18} and the complementary slackness conditions \eqref{3-19} and \eqref{3-23}, the KKT conditions of the problem minimizing $- h(\mathbf{w} | \mathbf{w}^{(\upsilon)})$ under the  constraints \eqref{3-12b} and \eqref{3-12c}  are given by
\begin{subequations} \label{app-the-1}
	\begin{align}
		- \nabla_{\mathbf{w}^*}  h (\mathbf{w}| \mathbf{w}^{o}) |_{\mathbf{w} = \mathbf{w}^o}  +  \tau   \mathbf{w}^o  + \mu  \nabla_{\mathbf{w}^*} h_1 ( \mathbf{w} | \mathbf{w}^o) |_{\mathbf{w} = \mathbf{w}^o}  & = 0,  \label{app-the-1a} \\ 
		\mu \left( h_1 ( \mathbf{w} | \mathbf{w}^o ) |_{\mathbf{w}  = \mathbf{w}^o} + \Omega  \right) & = 0,\label{app-the-1b} \\
		\tau ( \mathbf{w}^{o,H} \mathbf{w}^o - P_0) & = 0. \label{app-the-1c}
	\end{align}
\end{subequations}

Following from conditions (A1) and (A3) of the MM method, we have
\begin{subequations} \label{app-the-2}
	\begin{align}
		\nabla_{\mathbf{w}^*}  h (\mathbf{w}| \mathbf{w}^{o})  |_{\mathbf{w} = \mathbf{w}^o} & = 
		\nabla_{\mathbf{w}^*}  g (\mathbf{w} ) |_{\mathbf{w} = \mathbf{w}^o} , \label{app-the-2b} \\
		h_1 ( \mathbf{w}^o | \mathbf{w}^o) & = g_1 (\mathbf{w}^o) , \label{app-the-2c} \\
		\nabla_{\mathbf{w}^*} h_1 ( \mathbf{w} | \mathbf{w}^o) |_{\mathbf{w} = \mathbf{w}^o} & =  \nabla_{\mathbf{w}^*} g_1 (\mathbf{w}) |_{\mathbf{w} = \mathbf{w}^o}, \label{app-the-2d}
	\end{align}
\end{subequations}
where \eqref{app-the-2c} and  \eqref{app-the-2d} follows from the fact that the SCA method also satisfies conditions (A1) and (A3).
Substituting \eqref{app-the-2b}, \eqref{app-the-2c}, and \eqref{app-the-2d} into the KKT conditions \eqref{app-the-1}, we can obtain
\begin{subequations} \label{app-the-3}
	\begin{align}
		-\nabla_{\mathbf{w}^*}  g (\mathbf{w}) |_{\mathbf{w} = \mathbf{w}^o}  +  \tau   \mathbf{w}^o  + \mu  \nabla_{\mathbf{w}^*} g_1 (\mathbf{w}) |_{\mathbf{w} = \mathbf{w}^o} & = 0,  \label{app-the-3a} \\ 
		\mu  \left( g_1 (\mathbf{w}^o) + \Omega  \right)  & = 0,\label{app-the-3b} \\
		\tau ( \mathbf{w}^{o,H} \mathbf{w}^o - P_0) & = 0. \label{app-the-3c}
	\end{align}
\end{subequations}
Notice that the KKT conditions of Problem \eqref{2} for the case of a single communication user who is an extended interference, are comprised of three equations of \eqref{app-the-3}.

Hence, the proof is completed.
\bibliographystyle{IEEEtran}
\bibliography{ref}

\end{document}